\documentclass{aa}

\usepackage[T1]{fontenc}
\usepackage{amssymb}
\usepackage{graphicx}
\usepackage{amsmath}
\usepackage{bm}
\usepackage{booktabs}
\usepackage{longtable}
\usepackage{xcolor}
\usepackage{url}
\usepackage{natbib}
\usepackage{tikz}
\usetikzlibrary{arrows.meta,calc,positioning}
\usepackage{pgfplots}
\pgfplotsset{compat=1.18}
\usepgfplotslibrary{colormaps,groupplots}
\pgfplotsset{colormap={hamcor}{rgb255(0cm)=(255,255,255) rgb255(1cm)=(254,224,144)
  rgb255(2cm)=(253,141,60) rgb255(3cm)=(215,48,31) rgb255(4cm)=(127,0,0)}}
\pgfplotsset{colormap={hamdiv}{rgb255(0cm)=(33,102,172) rgb255(1cm)=(146,197,222)
  rgb255(2cm)=(247,247,247) rgb255(3cm)=(244,165,130) rgb255(4cm)=(178,24,43)}}

\newcommand{\rg}{r_{\rm g}}
\newcommand{\rgc}{r_{\rm g}/c}
\newcommand{\Ham}{\mathcal{H}}
\newcommand{\Hmag}{\mathcal{H}_{\rm mag}}
\newcommand{\Hlag}{\mathcal{H}_{\rm lag}}
\newcommand{\Hrad}{\mathcal{H}_{\rm rad}}
\newcommand{\Mbh}{M_{\rm BH}}
\newcommand{\Dil}{\mathcal{D}}
\newcommand{\Iobs}{I_{\rm obs}}
\newcommand{\Imax}{I_{\max}}
\newcommand{\nhat}{\hat{\bm n}_{\rm obs}}
\newcommand{\xmm}{\textit{XMM-Newton}}

\title{HAMCOR: inferring X-ray coronal geometry from reverberation lags with a Hamiltonian
formulation and exact photon transfer in Schwarzschild and Kerr spacetimes}

\author{F.~Buffoli\inst{1}}
\titlerunning{HAMCOR: coronal geometry from reverberation lags}
\authorrunning{F. Buffoli}

\institute{Universit\`a degli Studi di Brescia, Brescia, Italy\\
\email{fabio.buffoli@unibs.it}}

\date{}

\begin{document}

\abstract
{X-ray reverberation lags probe the geometry of the corona around accreting black holes,
but are usually interpreted with parametric geometries such as a lamppost.}
{We develop a geometry-agnostic method to infer the coronal emissivity distribution from
reverberation lags and test whether current data require an extended corona.}
{HAMCOR represents the corona as a non-negative emissivity field on a grid and minimises a
Hamiltonian combining a smoothness prior with lag likelihoods and the measured illumination
fraction. Photon transfer is computed by exact null-geodesic integration in Schwarzschild
and Kerr spacetimes, including the energy shifts of the reflected spectrum; the model passes
through the same cross-spectral estimator as the data.}
{Lag-frequency spectra constrain about four combinations of the emissivity; centroids are
recovered within $2$--$6\,\rg$. In eight \xmm{} observations of Mrk~335, 1H~0707$-$495,
IRAS~13224$-$3809 and MCG$-$6$-$30$-$15, six reflection fractions exceed the maximum
reachable around a non-spinning hole; at $a=0.998$ seven become reachable and require
compact coronae ($R_c\simeq2$--$3\,\rg$), in tension with the lags in three epochs of
Mrk~335. Lag-energy spectra fitted with the energy-resolved transfer quadruple the number
of constrained parameters. A lamppost allowed down to the horizon is preferred by the
Akaike criterion in all cases but one, Mrk~335 in 2006 at maximal spin, where it is
disfavoured only if $\sigma_I\lesssim0.03$, whereas alternative spectral models spread $I$
by $\pm0.1$.}
{Current data are consistent with, but do not require, an extended corona. Excluding a
lamppost requires a reflection fraction measured to a few per cent independently of the
spectral model together with a high-coherence lag-energy spectrum.}


\maketitle

\section{Introduction}
\label{sec:intro}

The hard X-ray continuum of active galactic nuclei (AGN) is produced by inverse-Compton
scattering of disc photons in a hot, compact corona \citep{Haardt1991,Haardt1993}. Its
geometry sets how the accretion power is dissipated close to the black hole and is still
debated: a hot inner flow \citep{Narayan1994}, a magnetically powered layer above the
disc \citep{Beloborodov1999,Merloni2003}, or the base of a jet \citep{Markoff2005} have
all been proposed.

X-ray reverberation offers a direct probe. Coronal photons reflected by the disc arrive
later than the direct continuum, producing soft lags at high Fourier frequencies, first
seen in 1H~0707$-$495 \citep{Fabian2009,Zoghbi2010} and now in dozens of AGN
\citep{DeMarco2013,Kara2016}; at lower frequencies hard lags from the propagation of
fluctuations in the flow dominate \citep{Kotov2001,Arevalo2006}. The amplitude of the
soft lag scales with black-hole mass \citep{DeMarco2013} and its frequency dependence
encodes the transfer function of the disc--corona system
\citep{Uttley2014,Cackett2014,Wilkins2016}.

Most interpretations use a lamppost, an isotropic point source on the spin axis
\citep{Dauser2013,Cackett2014,Chainakun2015}, which typically yields heights of a few
$\rg$. Spectral-timing and emissivity studies point instead to coronae that are radially
extended and variable \citep{Wilkins2012,WilkinsGallo2015,Wilkins2015}, and parametric
extensions (extended axial sources, two lampposts, disc-coronae) have been explored
\citep{Miniutti2004,Chainakun2019,Poutanen2018}.

This paper takes the complementary, non-parametric route. HAMCOR represents the corona as
an emissivity field on a grid and selects, among all fields compatible with the lags and
the measured reflection fraction, the smoothest one, by minimising a Hamiltonian
(Section~\ref{sec:method}). Compared with the first version of this work, the method has
been rebuilt around four requirements that any such inference must meet: (i) photon
transfer in curved spacetime computed exactly rather than perturbatively; (ii) the model
compared with the data through the same estimator, including dilution by the continuum
and an intrinsic continuum lag; (iii) the reflection fraction measured from the spectra
in the same spacetime as the timing model; (iv) an explicit account of how much of the
emissivity field the data can actually constrain. These choices change several earlier
conclusions, which we state explicitly where they occur. We then extend the transfer to
Kerr spacetime and to the energy shifts of the reflected spectrum, which turns the
lag-energy spectrum into a constraint (Sections~\ref{sec:spin} and~\ref{sec:eres-results}).

Section~\ref{sec:method} describes the method, Section~\ref{sec:data} the data reduction,
lag measurements and spectral fits, Section~\ref{sec:validation} the validation on
synthetic data, Section~\ref{sec:results} the results and Section~\ref{sec:discussion} the
comparison with lamppost models and the literature. Symbols are collected in
Appendix~\ref{app:nomenclature}. Lengths are in $\rg=G\Mbh/c^2$ and times in
$t_g=G\Mbh/c^3$.

\section{Method}
\label{sec:method}

\begin{figure}
\centering
\resizebox{\columnwidth}{!}{%
\begin{tikzpicture}[
  font=\footnotesize,
  box/.style={draw, rounded corners=2pt, align=center, minimum height=9mm, text width=28mm},
  data/.style={box, fill=blue!6},
  model/.style={box, fill=orange!12},
  opt/.style={box, fill=green!10, text width=42mm},
  res/.style={box, fill=gray!12},
  arr/.style={-{Stealth[length=2mm]}, thick},
  lab/.style={font=\scriptsize\itshape, text=black!60}]
\node[data] (evt) at (0,0)      {EPIC-pn events\\(SAS 22.1)};
\node[data] (lc)  at (-2.1,-1.4) {Light curves\\soft, hard, $E_k$};
\node[data] (sp)  at (2.1,-1.4)  {Spectrum\\+ RMF / ARF};
\node[data] (lag) at (-2.1,-2.8) {Lag--frequency\\$\tau_{\rm obs}(\nu_j)\pm\sigma_j$};
\node[data] (rf)  at (2.1,-2.8)  {\textsc{relxill} ($a=0$)\\$I_{\rm obs}$, $R_s$, $R_h$, $R_E$};
\node[model] (grid) at (6.0,-1.4) {Coronal grid\\$\bm w\in\Delta^{N-1}$};
\node[model] (gr)   at (6.0,-2.8) {Schwarzschild ray tracing\\$\Omega_{ik}$, $E_i$, $L_{ik}$};
\node[model] (psi)  at (6.0,-4.6) {$\tilde\psi(\nu;\bm w)$, $I(\bm w)$\\+ dilution, continuum lag};
\node[opt] (H)   at (0,-4.6) {$\mathcal{H}=\mathcal{H}_{\rm mag}+\mathcal{H}_{\rm lag}+\mathcal{H}_{\rm rad}$\\(MAP objective)};
\node[opt] (pgd) at (0,-6.0) {Projected gradient, Armijo\\multistart, bootstrap};
\node[res] (chk) at (-2.1,-7.5) {Lamppost fit\\lag--energy check};
\node[res] (map) at (2.1,-7.5)  {$w(R,z)$, centroid\\uncertainties};
\draw[arr] (evt) -| (lc);   \draw[arr] (evt) -| (sp);
\draw[arr] (lc) -- (lag);   \draw[arr] (sp) -- (rf);
\draw[arr] (grid) -- (gr);  \draw[arr] (gr) -- (psi);
\draw[arr] (lag) |- ($(H.north west)!0.25!(H.north east)+(0,3mm)$) -- ($(H.north west)!0.25!(H.north east)$);
\draw[arr] (rf)  |- ($(H.north west)!0.75!(H.north east)+(0,3mm)$) -- ($(H.north west)!0.75!(H.north east)$);
\draw[arr] (psi.west) -- (H.east);
\draw[arr] (H) -- (pgd);
\draw[arr] (pgd.south) -- ++(0,-2.5mm) -| (chk); \draw[arr] (pgd.south) -- ++(0,-2.5mm) -| (map);
\draw[arr, dashed] (lc.west) -- ++(-0.35,0) |- (chk.west)
      node[lab, pos=0.25, left, align=right] {$\tau(E)$\\not fitted};
\node[lab, above=0mm of grid] {forward model};
\node[lab, above=0mm of evt, yshift=0mm] {};
\end{tikzpicture}%
}
\caption{HAMCOR workflow. Light curves and spectra are extracted with \textsc{sas}; lags and the reflection fraction and dilution factors are measured from them and compared with the forward model (exact Schwarzschild transfer) through the Hamiltonian. The lag-energy spectrum is used only as an independent check.}
\label{fig:flowchart}
\end{figure}

\subsection{Coronal and disc representation}
\label{sec:repr}

The corona is described by $N=n_Rn_zn_\phi=10\times10\times8=800$ cells with
cylindrical coordinates $(R_i,z_i,\phi_i)$, logarithmically spaced in
$3\le R\le30\,\rg$ and $0.1\le z\le20\,\rg$ and uniform in $\phi$. Each cell carries an
emissivity weight $w_i\ge0$ with $\sum_iw_i=1$, i.e. $\bm w$ lies on the probability
simplex $\Delta^{N-1}$; cells emit isotropically in their static frame. The disc is
geometrically thin, in the equatorial plane between $r_{\rm in}=6\,\rg$ (the ISCO of a
Schwarzschild hole) and $100\,\rg$, divided into $30$ logarithmic rings and $18$ azimuthal
sectors (540 elements). The observer lies at inclination $\theta_{\rm obs}$ in the
$x$--$z$ plane, $\nhat=(\sin\theta_{\rm obs},0,\cos\theta_{\rm obs})$
(Fig.~\ref{fig:flowchart}).

\subsection{Impulse response and transfer function}
\label{sec:transfer}

Let $\Omega_{ik}$ be the solid angle (in the frame of cell $i$) of the photons from cell
$i$ that reach disc element $k$, $E_i$ the solid angle of those that escape to infinity,
and $L_{ik}$ the extra arrival time at the observer of a photon reflected by element $k$
relative to one received directly from cell $i$. We define the impulse response of the
reflected flux to a $\delta$-function flare of the continuum
\citep{Blandford1982,Uttley2014},
\begin{equation}
  \psi(t;\bm w)=\frac{\sum_i w_i\sum_k \Omega_{ik}\,\delta(t-L_{ik})}{\sum_i w_i a_i},
  \qquad a_i=\sum_k\Omega_{ik},
  \label{eq:psi}
\end{equation}
normalised to $\int\psi\,{\rm d}t=1$, and its Fourier transform, the transfer function,
\begin{equation}
  \tilde\psi(\nu;\bm w)=\frac{\sum_i w_i G_i(\nu)}{\sum_i w_ia_i},\qquad
  G_i(\nu)=\sum_k\Omega_{ik}\,e^{-2\pi{\rm i}\nu L_{ik}} .
  \label{eq:psitilde}
\end{equation}
Reflected photons are counted with their band-integrated rates (Section~\ref{sec:dilution});
energy shifts of the reflected spectrum and the angular emissivity of the disc are not
modelled (Section~\ref{sec:limitations}). The first moment
$\langle t\rangle=\int t\,\psi\,{\rm d}t$ is the response-weighted mean delay.

\subsection{Photon transfer in Schwarzschild spacetime}
\label{sec:gr}

$\Omega_{ik}$, $E_i$ and $L_{ik}$ are obtained by integrating null geodesics. In units
$G=c=M=1$ a geodesic with impact parameter $b$ lies in a plane through the hole and obeys
${\rm d}^2u/{\rm d}\varphi^2+u=3u^2$ with $u=1/r$. The same orbit is generated by the
central force
\begin{equation}
  \frac{{\rm d}^2\bm x}{{\rm d}\lambda^2}=-\frac{3h^2\,\bm x}{r^5},\qquad
  h=\left|\bm x\times\frac{{\rm d}\bm x}{{\rm d}\lambda}\right| ,
  \label{eq:geodesic}
\end{equation}
(the Binet equation with $F\propto r^{-4}$), which we integrate in three dimensions with a
fourth-order Runge--Kutta scheme and step ${\rm d}\lambda=0.01\,r$. Choosing the initial
speed such that $h=b$ and ${\rm d}r/{\rm d}\lambda=\pm[1-b^2(1-2/r)/r^2]^{1/2}$, the
coordinate time follows exactly from ${\rm d}t/{\rm d}\varphi=r^2/[b(1-2/r)]$ and
${\rm d}\varphi/{\rm d}\lambda=h/r^2$,
\begin{equation}
  \frac{{\rm d}t}{{\rm d}\lambda}=\left(1-\frac{2}{r}\right)^{-1},
  \label{eq:dtdl}
\end{equation}
also for radial rays. A photon emitted with local direction $(n_r,n_t)$ in the static frame
at radius $r_0$ has $b=r_0n_t(1-2/r_0)^{-1/2}$.

For every $(R,z)$ of the grid we shoot $4\times10^4$ rays on a Fibonacci lattice of local
directions. A ray is assigned to the disc element where it first crosses the equatorial
plane at $6\le r\le100$, is captured if $r<2.05$, and escapes if $r>600$; rays crossing the
plane inside $r_{\rm in}$ or beyond the outer edge continue. Then
$\Omega_{ik}=4\pi N_{ik}/N_{\rm rays}$, $E_i=4\pi N_{{\rm esc},i}/N_{\rm rays}$, and the
mean coordinate time $\langle t_{\rm SP}\rangle_{ik}$ of the rays reaching $k$ is stored.
Azimuthal symmetry gives the other seven $\phi$ cells by rotation. The arrival-time offset
of a photon leaving point $\bm x$ towards a distant observer,
\begin{equation}
  T(\bm x)=\lim_{r\to\infty}\left[t-\bm x_\infty\cdot\nhat-2\ln r\right],
  \label{eq:T}
\end{equation}
is found by bisection on the emission angle in the plane of $\bm x$ and $\nhat$, measuring
the asymptotic direction as a signed angle so that rays deflected by more than $\pi$ are
handled correctly. The lag matrix is
\begin{equation}
  L_{ik}=\langle t_{\rm SP}\rangle_{ik}+T(\bm x_k)-T(\bm x_i)\ \ge0 .
  \label{eq:lagmatrix}
\end{equation}
In flat space $T=-\bm x\cdot\nhat$ and equation~(\ref{eq:lagmatrix}) reduces to the
Euclidean path difference.

The integrator reproduces exactly the flat-space solid angles, distances and delays when
the force is switched off, the analytic capture cone of a static emitter
($\sin\psi_c=3\sqrt3\,r^{-1}(1-2/r)^{1/2}$; captured fraction $0.2034$ at $r=5$, analytic
$0.2034$) and the weak-field limit at large distance; every reflected ray arrives after
the direct one, also for sources behind the hole (Appendix~\ref{app:tests}). Shapiro
delay, light bending, capture and occultation are thus all included exactly. The
first-order (weak-field) Shapiro treatment of the first version of this work, its
correction, and the comparison with the exact transfer are given in Appendix~\ref{app:weak}.
The tables are computed once per grid (about 10 minutes on one core, plus about 2 minutes per
inclination for $T$) and cached.

\subsection{Kerr transfer}
\label{sec:kerr}

For a spinning hole we integrate null geodesics of the Kerr metric in Kerr--Schild
Cartesian coordinates, regular at the horizon, as the Hamiltonian flow
$\dot x^\mu=\partial\mathcal{K}/\partial p_\mu$, $\dot p_\mu=-\partial\mathcal{K}/\partial x^\mu$
with $\mathcal{K}=\frac12g^{\mu\nu}p_\mu p_\nu$ (fourth-order Runge--Kutta, step proportional
to $r$, spatial derivatives of $g^{\mu\nu}$ by central differences). Cells emit isotropically
in the frame of a static observer; cells within $g_{tt}>-0.02$ of the ergosphere are masked.
The disc extends from the prograde ISCO ($1.24\,\rg$ at $a=0.998$) to $100\,\rg$. Orbits are
not planar, but the spacetime is axisymmetric, so $\Omega_{ik}$, $E_i$ and $L_{ik}$ are
computed for one azimuthal column of cells and rotated. The observer delay $T(\bm x)$ of
each cell is obtained by shooting: a Newton iteration with backtracking on a
two-dimensional (exponential-map) parametrisation of the local emission direction,
started from several directions for points behind the hole, keeping the earliest
converged arrival (the primary image). For the disc, $T$ and the redshift $g$ of a
prograde Keplerian flow are obtained by backward tracing from the observer's image plane.
The offset between Kerr--Schild and Boyer--Lindquist time depends on $r$ only and cancels
in $L_{ik}$. At $a=0$ the module reproduces the Schwarzschild transfer: identical
$\Omega_{ik}$, $E_i$ and capture fractions, $\Imax=0.561$ against $0.564$, and the
$\chi^2_{\rm lag}$ of all eight observations within $0.1$ (Table~\ref{tab:kerr}); the parity
response of Section~\ref{sec:parity}, which must vanish at $a=0$, is $\le10^{-2}$, which we
take as the numerical floor. The Kerr fits use a grid extended inwards to $R=1\,\rg$
($10\times10\times8$ cells); the lamppost is scanned from $h=1.12\,\rg$, just outside the
horizon at $a=0.998$, to $40\,\rg$.

\subsection{Illumination fraction}
\label{sec:illum}

The relativistic reflection fraction of \textsc{relxill} is defined as the ratio of
photons reaching the disc to photons reaching infinity \citep{Dauser2016}. The model
equivalent, with captured photons excluded, is
\begin{equation}
  I(\bm w)=\frac{\sum_iw_ia_i}{\sum_iw_i(a_i+E_i)}=\frac{R_f}{1+R_f} .
  \label{eq:illum}
\end{equation}
Fig.~\ref{fig:illum} shows $I$ for an on-axis source and for single cells. Light bending
raises the illumination of sources close to the hole, but capture removes photons from
both the reflected and the escaping budget, and a disc truncated at $6\,\rg$ cannot
intercept more than $\Imax=0.564$ of the useful photons for any cell of the grid
($R_f\le1.29$); a lamppost reaches at most $R_f\simeq0.9$, in agreement with
\citet{Dauser2014}. The solid-angle formula $\cos\vartheta\,{\rm d}A/d^2$ used in the
first version over-counts the disc seen from cells close to it and gave $I$ up to $0.74$.

\subsection{Band lags and dilution}
\label{sec:dilution}

The soft and hard bands contain the direct continuum and the reflected component in the
count-rate ratios $R_s$ and $R_h$ (reflected over continuum, measured from the spectrum;
Section~\ref{sec:spectra}). With a common continuum light curve $c(t)$, the band light
curves are $s=c\ast(\delta+R_s\psi)$ and $h=c\ast(\delta+R_h\psi)$, so the lag of the soft
band with respect to the hard band is
\begin{equation}
  \tau(\nu)=\frac{\arg[1+R_s\tilde\psi(\nu)]-\arg[1+R_h\tilde\psi(\nu)]}{2\pi\nu}
  \label{eq:tau}
\end{equation}
\citep[cf.][]{Uttley2014,Wilkins2016}, with negative values for soft lags. In the
low-frequency limit
\begin{equation}
  \tau\to-\Dil\,\langle t\rangle,\qquad \Dil=\frac{R_s}{1+R_s}-\frac{R_h}{1+R_h},
  \label{eq:lowf}
\end{equation}
which shows explicitly how the measured soft--hard lag is a diluted proxy of the
reverberation delay: for the values of Table~\ref{tab:spectral} only
$\Dil=0.16$--$0.40$ of the mean delay appears in the lag. Dilution is therefore not a
correction applied to the data but part of the model, with $R_s$, $R_h$ and their
uncertainty propagated into the likelihood (Section~\ref{sec:hamiltonian}).

The propagation of fluctuations through the flow produces an intrinsic hard lag of the
continuum \citep{Kotov2001,Arevalo2006}, which dominates at low frequencies and, in
low-flux states, can reach the reverberation band. We add
\begin{equation}
  \tau_{\rm mod}(\nu)=\tau(\nu)+A\left(\nu/\nu_0\right)^{-\gamma_c},
  \label{eq:cont}
\end{equation}
with $A\ge0$, $\gamma_c\in\{0,0.5,1,1.5,2\}$ and $\nu_0$ the geometric mean of the fitted frequencies. $(A,\gamma_c)$ are profiled
analytically at every evaluation (non-negative weighted least squares in $A$ for each $\gamma_c$)
and counted in the degrees of freedom; by the envelope theorem the gradient with respect to
$\bm w$ is that at the optimal $(A,\gamma_c)$.

\subsection{Energy-resolved transfer}
\label{sec:eres}

Equation~(\ref{eq:tau}) treats the reflected light in each band as one component with a
single transfer function. Most of the energy structure of lag-energy spectra, however,
comes from the gravitational and Doppler shifts of the reflected spectrum
\citep{Cackett2014,Ingram2019}. Since the photon-number intensity per unit energy
transforms as $N_{\rm obs}(E)=g^2N_{\rm em}(E/g)$, the reflected photons that disc element
$k$ contributes to a channel $[E_1,E_2]$ are, per illuminating photon,
\begin{equation}
  S_{kE}=\frac{\Omega^{\rm sky}_k}{A_k}\,g_k^3\int_{E_1/g_k}^{E_2/g_k}N_{\rm refl}(E')\,{\rm d}E' ,
  \label{eq:SkE}
\end{equation}
where $N_{\rm refl}$ is the rest-frame reflection spectrum (\textsc{xillver} or
\textsc{xillverCp}, \citealt{Garcia2013}, at the best-fitting $\Gamma$, $A_{\rm Fe}$, $\xi$ and,
where fitted, density), $g_k$ and the solid angle $\Omega^{\rm sky}_k$ that the element
subtends on the observer's sky (lensing and projection included) come from the image-plane
tracing, and $A_k$ is the element area. The transfer function of channel $E$ is
\begin{equation}
  \tilde\psi_E(\nu)=\frac{\sum_iw_i\sum_k\Omega_{ik}S_{kE}\,e^{-2\pi{\rm i}\nu L_{ik}}}
  {\sum_iw_i\sum_k\Omega_{ik}S_{kE}} ,
  \label{eq:psiE}
\end{equation}
and the reference band (all other channels) responds with the average of $\tilde\psi_{E'}$
weighted by the reflected count rates. The channel amplitudes $R_E$ are those of the
spectral fit, so the energy shifts change the time structure of the response in each
channel, not its reflected fraction; for $S_{kE}$ independent of $E$ the model reduces to
the band-integrated one of Appendix~\ref{app:lagenergy}. The lag-energy likelihood
$\mathcal{H}_{\rm le}=\frac12\min_c\sum_E[(\tau_E+c-\tau_{{\rm obs},E})/\sigma_E]^2$, with a
profiled additive offset $c$, has an analytic gradient and is added to $\Ham$ for the joint
fits of Section~\ref{sec:eres-results}. The energy shift of the illuminating photons and
the angular dependence of the reflected emission beyond the template inclination are
neglected. In Schwarzschild, $g_k$ and $\Omega^{\rm sky}_k$ are taken from the Kerr module
at $a=0$.

\subsection{Estimator-matched likelihood}
\label{sec:likelihood}

Lags are measured by averaging cross-spectra over frequency bins $j$, each containing
Fourier frequencies $\nu_k$ weighted by the power $p_k$ (Section~\ref{sec:lags}). The model
is passed through the same estimator,
\begin{align}
  Z_j&=\sum_{k\in j}p_k\,[1+R_h\tilde\psi(\nu_k)]^*[1+R_s\tilde\psi(\nu_k)], \nonumber\\
  \tau_j&=\frac{\arg Z_j}{2\pi\bar\nu_j},
  \label{eq:estimator}
\end{align}
which removes the bias of comparing a binned, power-weighted phase with the model at the
bin centre. The lag likelihood is
\begin{align}
  \Hlag&=\frac{\alpha}{2}\sum_j\frac{(\tau_{{\rm mod},j}-\tau_{{\rm obs},j})^2}{\sigma_j^2}, \nonumber\\
  \sigma_j^2&=\sigma_{\tau,j}^2+\left(\tau_{{\rm obs},j}\,\sigma_\Dil/\Dil\right)^2,
  \label{eq:hlag}
\end{align}
where the second term propagates the dilution uncertainty. The illumination term is
\begin{equation}
  \Hrad=\frac{\beta}{2}\frac{\left[I(\bm w)-\Iobs\right]^2}{\sigma_I^2} .
  \label{eq:hrad}
\end{equation}
With $\alpha=\beta=1$, $2(\Hlag+\Hrad)$ is the $\chi^2$ of the data.

\subsection{The Hamiltonian}
\label{sec:hamiltonian}

The emissivity is the minimum of
\begin{align}
  \Ham(\bm w)&=\Hmag(\bm w)+\Hlag(\bm w)+\Hrad(\bm w), \nonumber\\
  \Hmag&=\frac{J\kappa}{2}\sum_{\langle i,j\rangle}(w_i-w_j)^2 ,
  \label{eq:ham}
\end{align}
over $\Delta^{N-1}$, where the sum runs over nearest-neighbour pairs (radial, vertical and
cyclic azimuthal) and $\kappa=N^2/N_{\rm pairs}$ makes $J$ independent of the grid size.
$\Hmag$ is a Gaussian Markov random-field prior \citep{Rue2005} that penalises emissivity
jumps between adjacent cells; $J$ sets its strength, and the minimum of $\Ham$ is the maximum
a posteriori (MAP) solution. We keep the subscript of the first version, where the term was
written as a magnetic coupling $-J\sum w_iw_j\,\hat{\bm n}_i\cdot\hat{\bm n}_j$; that form is
unbounded from below on the simplex and drives solutions towards two-cell clumps, and the
orientations it contains are not constrained by timing data. The pair-production and energy-budget
barriers of the first version were never active for the fitted sources and are omitted.

The first version compared a binned impulse response with a Gaussian ``observed''
response centred on the measured lag. That construction is not a likelihood of the data
and its shape is arbitrary; it is replaced by equation~(\ref{eq:hlag}). We retain the
earlier forms only to test the sensitivity of the recovery to this choice
(Gaussian, top-hat and uniform $\psi_{\rm obs}$; Section~\ref{sec:robustness}).

\subsection{Optimisation and uncertainties}
\label{sec:optim}

$\Ham$ is minimised by projected gradient descent on the simplex (Euclidean projection,
\citealt{Duchi2008}) with Armijo backtracking along the projection arc
\citep{Bertsekas1999}. Analytic gradients of all terms are verified against finite
differences. A continuation in the weight of the data terms over the first 300 iterations
avoids early trapping; we run 16 starts (uniform and random Dirichlet) for 3000 iterations
and keep the lowest $\Ham$ (8 starts in the synthetic tests). Statistical uncertainties come from a bootstrap: 30 refits after
resampling the lags within their errors, $\Iobs$ within $\sigma_I$ and $\Dil$ within its
error. Systematics are the centroid shifts under flat and weak-field spacetime, $J=0.3$ and
$3$, and $\Mbh\pm1\sigma$. We summarise solutions by the emissivity centroid
$(R_c,z_c)=\sum_iw_i(R_i,z_i)$. The full analysis of one source (16-start baseline, seven variants, 30 bootstrap refits) takes
1.5--4 minutes on two cores once the transfer tables exist.

\subsection{Information content}
\label{sec:info}

With four to seven lag points and one illumination fraction the problem is
underdetermined by two orders of magnitude. We quantify the constrained directions with the
Fisher matrix of the data terms at the solution projected on the tangent space of the
simplex, $F=P^{\sf T}\mathsf{J}^{\sf T}\Sigma^{-1}\mathsf{J}P$, where $\mathsf J$ is the
Jacobian of the model observables. The number of effective parameters,
$N_{\rm eff}=\sum_n\lambda_n/(\lambda_n+\lambda_{\rm prior})$ with the eigenvalues
$\lambda_n$ of $F$ and $\lambda_{\rm prior}$ the curvature of $\Hmag$, counts the
combinations of $\bm w$ that the data determine; everything else is set by the prior. The corresponding eigenvectors show which combinations these are; we project them on
$(R,z)$ and correlate them with physical descriptors of each cell (mean delay
$\langle t\rangle_i=\sum_k\Omega_{ik}L_{ik}/a_i$, radius, height, illumination) in
Section~\ref{sec:fits} (Fig.~\ref{fig:modes}).

\section{Data}
\label{sec:data}

\subsection{Observations and reduction}
\label{sec:obs}

We use the eight EPIC-pn \citep{Struder2001} observations of Table~\ref{tab:obs}: five
epochs of Mrk~335 between 2006 and 2019 covering a factor $\sim40$ in soft count rate, and
one long observation each of 1H~0707$-$495, IRAS~13224$-$3809 and MCG$-$6$-$30$-$15. The
Cyg~X-1 data used in the first version are not included: the observation analysed there
is not available in the \xmm{} archive under the quoted identifier, and a meaningful
comparison would require timing-mode data reduced with a dedicated procedure.

The data were reduced with \textsc{sas} 22.1.0 \citep{Gabriel2004} and the current
calibration files: \texttt{cifbuild}, \texttt{odfingest}, \texttt{epproc}; single and
double events (\texttt{PATTERN}$\le4$, \texttt{FLAG}$=$0); background flares were
identified from the 10--12\,keV single-event light curve (threshold $0.5$\,ct\,s$^{-1}$) and excluded from the spectra; the timing analysis uses the full exposure with the background subtracted, so that segments are not broken by short flares. Source counts
were extracted from a $40''$ circle and background from a $50''$ circle on the same CCD;
the background circle is placed automatically where it lies entirely within the active
window (at least 98 per cent of its area containing events), with the radius reduced
where needed in Small Window mode. Pile-up is negligible: \texttt{epatplot} gives
observed-to-model pattern fractions within 2 per cent of unity. Light curves in
100\,s bins were extracted in 0.3--1, 1.5--4\,keV and 16 energy channels from 0.3 to
10\,keV and corrected with \texttt{epiclccorr}; bins with fractional exposure below 0.9 of
the median were discarded and gaps of up to five bins (at most 0.8 per cent of the data)
linearly interpolated. Spectra were grouped to a minimum of 25 counts and an
oversampling of 3 with \texttt{specgroup}, with responses from \texttt{rmfgen} and
\texttt{arfgen}.

\subsection{Lag-frequency spectra}
\label{sec:lags}

Light curves are divided into contiguous segments of 25.6\,ks (12.8\,ks for Mrk~335 in
2009 and 6.4\,ks for 1H~0707$-$495 and MCG$-$6$-$30$-$15, where gaps or flares break the
series); the series is end-matched within each
segment \citep{Fougere1985} to remove the red-noise leak, which otherwise biases the
lowest-frequency lags by $\sim15$ per cent in our simulations. Cross-spectra are averaged
over segments and in logarithmic frequency bins; lag errors follow
\citet{Bendat2010} with the coherence corrected for Poisson noise
\citep{Vaughan1997,Nowak1999}. Segments are independent (no overlap), so that the
number of averaged cross-spectra entering the errors is the true number of independent
estimates. The fitted frequency band is chosen for each source where reverberation lags
are expected and measured \citep{DeMarco2013,Kara2016}
(Table~\ref{tab:lags}, Fig.~\ref{fig:lagfreq}).

\subsection{Spectral fits and the reflection fraction}
\label{sec:spectra}

$\Iobs$, $R_s$ and $R_h$ are measured from the time-averaged pn spectra with
\textsc{xspec} 12.15.1 \citep{Arnaud1996} (\textsc{heasoft} 6.36), abundances of
\citet{Wilms2000} and cross-sections of \citet{Verner1996}, over 0.3--10\,keV. The model is
\begin{equation*}
  \textsc{tbabs}\times[\textsc{zxipcf}]\times(\textsc{relxill}+\textsc{xillver}\,[+\textsc{zgauss}]),
\end{equation*}
with \textsc{relxill} 2.9.0 \citep{Garcia2014,Dauser2016}: Galactic absorption fixed;
partial coverers for the low-flux epochs of Mrk~335 (two for 2018 and 2019) and two for
the warm absorber of MCG$-$6$-$30$-$15; a distant, neutral, reflection-only
\textsc{xillver} component with $\Gamma$ and $A_{\rm Fe}$ tied to the relativistic one; a
narrow Fe\,\textsc{xxvi} line at 6.97\,keV for Mrk~335. For 1H~0707$-$495 and
IRAS~13224$-$3809 the relativistic component is \textsc{relxillCp} with the disc density
free in $15\le\log n_{\rm e}\le20$ \citep{Garcia2016}. A systematic error of 1 per cent is
added to account for calibration uncertainties of spectra with $10^5$--$10^6$ counts; $\chi^2$
without it is also reported.

Spin and inner radius are chosen for consistency with the timing model, not to optimise
the spectral fit: $a=0$ and $R_{\rm in}=R_{\rm ISCO}=6\,\rg$, the same disc as in
HAMCOR. With $R_{\rm br}<6\,\rg$ a broken emissivity is then unconstrained, so a single
index $q$ is used. Because the spectra of several sources prefer a rapidly spinning hole,
each fit is repeated at $a=0.998$, and $|I(0.998)-I(0)|$ and $|\Dil(0.998)-\Dil(0)|$ are
added in quadrature to the uncertainties as a spin systematic. The statistical errors are
the larger of the scatter over 100 parameter sets drawn from the fit covariance and the
1$\sigma$ interval of a $\chi^2$ profile in $R_f$ in which all other parameters are refitted
(the \textsc{xspec} \texttt{error} command fails on the $q$--$R_f$ degeneracy; the profile
also restarts the fit whenever it finds a lower minimum).

$R_s$ and $R_h$ are the ratios of the folded count rates of the variable reflected and
continuum components in the two bands, obtained by setting the reflection fraction of
\textsc{relxill} to zero (continuum only) and to $-R_f$ (reflection only). The distant
reflector and the narrow line respond on time-scales much longer than those of
Table~\ref{tab:lags} and contribute only to the constant part of the light curves; they
are excluded from $R_{s,h}$. The same ratios in the 16 lag-energy channels enter the
lag-energy check (Appendix~\ref{app:lagenergy}).

\begin{table*}
\centering
\caption{Reflection fits (\textsc{xspec} 12.15.1, \textsc{relxill} 2.9.0, $a=0$, $R_{\rm in}=R_{\rm ISCO}=6\,r_g$, single emissivity index, 1 per cent systematic; $\chi^2$ without the systematic in brackets). $R_f$: reflection fraction with 1$\sigma$ profile interval; $I_{\rm obs}=R_f/(1+R_f)$, its error combines covariance draws, the $R_f$ profile and the spin systematic $|I_{0.998}-I|$ ($I_{0.998}$: same model refitted at $a=0.998$). The error on $\Dil$ adds in quadrature the spin systematic and an emissivity-model systematic (difference between broken and single emissivity at $a=0.998$). $R_{s,h}$: reflected-to-continuum count-rate ratios of the variable components in the soft and hard bands; $\mathcal{D}=R_s/(1+R_s)-R_h/(1+R_h)$.}
\label{tab:spectral}
\small
\begin{tabular}{lcccccccc}
\toprule
Source & $\chi^2/\nu$ (stat.) & $R_f$ & $I_{\rm obs}$ & $I_{0.998}$ & $R_s$ & $R_h$ & $\mathcal{D}$ & $\mathcal{D}_{0.998}$ \\
\midrule
Mrk 335 (2006) & 496/169 (2132) & $3.73^{+0.15}_{-0.13}$ & $0.79\pm0.02$ & 0.81 & 1.19 & 0.63 & $0.16\pm0.10$ & 0.08 \\
Mrk 335 (2009) & 269/160 (330) & $1.85^{+0.15}_{-0.15}$ & $0.65\pm0.21$ & 0.86 & 4.70 & 0.77 & $0.39\pm0.15$ & 0.24 \\
Mrk 335 (2015) & 220/148 (229) & $2.52^{+0.18}_{-0.19}$ & $0.72\pm0.01$ & 0.72 & 5.40 & 0.83 & $0.39\pm0.13$ & 0.39 \\
Mrk 335 (2018) & 1088/138 (1194) & $7.16^{+0.64}_{-0.17}$ & $0.88\pm0.01$ & 0.88 & 3.68 & 0.89 & $0.31\pm0.07$ & 0.31 \\
Mrk 335 (2019) & 748/135 (787) & $10.48^{+0.27}_{-0.45}$ & $0.91\pm0.04$ & 0.95 & 3.14 & 0.93 & $0.28\pm0.03$ & 0.27 \\
1H 0707$-$495 & 879/147 (1667) & $10.51^{+0.30}_{-0.32}$ & $0.91\pm0.02$ & 0.93 & 5.69 & 0.81 & $0.40\pm0.02$ & 0.38 \\
IRAS 13224$-$3809 & 488/109 (618) & $25.45^{+0.68}_{-0.71}$ & $0.96\pm0.02$ & 0.98 & 4.58 & 0.84 & $0.36\pm0.02$ & 0.35 \\
MCG$-$6$-$30$-$15 & 309/165 (1112) & $0.71^{+0.05}_{-0.04}$ & $0.42\pm0.03$ & 0.45 & 4.17 & 1.33 & $0.24\pm0.14$ & 0.23 \\
\bottomrule
\end{tabular}
\end{table*}

The fits (Table~\ref{tab:spectral}, Fig.~\ref{fig:spectra}) are acceptable for Mrk~335 in
2006--2015 and for MCG$-$6$-$30$-$15 ($\chi^2_\nu=1.5$--$2.9$ with the systematic), and poor for
the low-flux states of Mrk~335 in 2018--2019 and for the two narrow-line Seyfert~1s
($\chi^2_\nu=4.5$--$8$). Their residuals are dominated by narrow features below 1\,keV, from
photoionised gas in the low states of Mrk~335 \citep{Grupe2008,Parker2019} and by
outflow absorption in the NLS1s, that the pn resolution does not separate. $\Dil$, the quantity entering the lag model, depends less on these choices than $R_f$ does,
but not negligibly: the spin changes it by $0.08$ and $0.15$ for Mrk~335 in 2006 and 2009
(included in the errors), and the broken-emissivity, high-spin fits of our preliminary analysis
differ from the final ones by up to $0.14$. We include this emissivity-model systematic, measured for each observation as the difference
between broken and single emissivity at $a=0.998$ (0.005--0.14), in quadrature in $\sigma_\Dil$
(Table~\ref{tab:spectral}); through equation~(\ref{eq:hlag}) it enters the likelihood, the
bootstrap and all centroids of Table~\ref{tab:fits}.

\section{Validation on synthetic data}
\label{sec:valsum}

Appendix~\ref{sec:validation} tests HAMCOR on three synthetic geometries (a compact on-axis
source, a column and a torus) with the exact transfer, realistic noise and the frequency
coverage of the data. The lags are reproduced to 5--9 per cent and the centroids to
$2$--$4\,\rg$ ($6\,\rg$ for the torus); with $N_{\rm eff}=4$ the rest of the map is the
smoothest field compatible with the data ($\rho_{3D}=0.2$--$0.5$). Fitting with flat-space
transfer displaces the centroid by $5$--$9\,\rg$ although the mean delay changes by only
12--14 per cent, so exact transfer is necessary for the inferred geometry, not only for the
lag amplitude. The results are stable for $0.1\le J\le10$; we adopt $J=1$. The part of the
emissivity that is odd under reflection about the observer's meridian plane is exactly
unobservable with lags and reflection fractions in Schwarzschild spacetime
(Appendix~\ref{sec:parity}).

\section{Results}
\label{sec:results}

\subsection{The illumination limit}
\label{sec:illum-limit}

Six of the eight observations have $\Iobs-2\sigma_I>\Imax$ (Table~\ref{tab:spectral}):
reflection fractions between $2.5$ and $25$, against a maximum of $1.3$ for any isotropic
coronal configuration around a non-spinning hole with the disc at the ISCO. Fitting them
with $\Hrad$ is possible but meaningless: the solution collapses onto the cells of maximum
illumination, just above the inner disc, with $\chi^2_I=80$--$1400$. This is the timing
counterpart of the spectral argument of \citet{Dauser2014}: large reflection fractions
require light bending stronger than Schwarzschild allows, i.e. a rapidly spinning hole, an
anisotropic or outflowing source, or additional reflection not described by the model
(e.g. returning radiation). For these six observations the baseline is fitted to the lags
alone and the illumination-constrained solution is reported only as a variant; the
illumination constraint is used for Mrk~335 in 2009 and for MCG$-$6$-$30$-$15. We regard
this as a limitation of a Schwarzschild forward model, not as evidence against the
reflection interpretation.

\subsection{Fits}
\label{sec:fits}

The fits are summarised in Table~\ref{tab:fits} and Figs.~\ref{fig:lagfreq} and
\ref{fig:maps}. All lag spectra are described within the errors
($\chi^2_{\rm lag}/\nu\le1.9$) except Mrk~335 in 2015 ($\chi^2=5.9$ for 2 dof, $p=0.05$), whose
lag of $-563\pm372$\,s in the lowest bins requires a mean delay ($39\,\rgc$) at the edge of what
the grid can produce with the measured dilution. The continuum-lag amplitude is non-zero in
six of eight fits but dominant only in the 2019 epoch of Mrk~335 ($A\simeq1200$\,s), whose band
lag is positive. All solutions
lie inside the grid and the bootstrap intervals are well defined.

The recovered coronae are extended: emissivity centroids at $R_c\simeq6$--$12\,\rg$ and
$z_c\simeq3$--$7\,\rg$, mean delays $\langle t\rangle\simeq18$--$31\,\rgc$, and maps
(Fig.~\ref{fig:maps}) with a broad distribution over the inner $\sim15\,\rg$ rather than a
compact source on the axis. The systematic shifts (spacetime, $J$, mass) are $1$--$4\,\rg$, comparable to the bootstrap
intervals, except for MCG$-$6$-$30$-$15, where the flat-space fit moves the centroid outwards by
$5\,\rg$. Following Section~\ref{sec:recovery}, the statements that the data support are the centroid
and the mean delay; the detailed shape of the maps is set largely by the prior, which is why
Fig.~\ref{fig:maps} shows bootstrap averages and Fig.~\ref{fig:mapscv} their scatter.

Fig.~\ref{fig:modes} shows what the data do constrain. The leading Fisher mode is, in seven of
eight observations, essentially the mean delay of the cells (correlation
$|\rho|=0.5$--$0.9$ with $\langle t\rangle_i$, and $0.4$--$0.7$ with the height): the lags fix
how much emission sits at long or short delays, i.e. mostly how high the corona is. Where the
illumination fraction is used it adds information that tracks $I_i$ (for
MCG$-$6$-$30$-$15 the two leading modes correlate with $I_i$ at $-0.83$ and $-0.55$, mixed
with the mean delay; $-0.83$ for the fourth mode of Mrk~335 in 2009). The remaining two or three
modes are weaker by one to three orders of magnitude and are not reducible to a single
descriptor; they encode the shape of the delay distribution sampled by the few frequency
bins.

\begin{table*}
\centering
\caption{HAMCOR fits (exact Schwarzschild transfer, $J=1$). $\nu$ counts the two continuum-lag parameters; $A$, $\gamma_c$: profiled continuum lag $A(\nu/\nu_0)^{-\gamma_c}$. Centroids: bootstrap median and 16--84 percentiles (30 resamplings of lags, $I_{\rm obs}$ and dilution). Syst.: largest centroid shift over flat space, weak-field Schwarzschild, $J=0.3$ and $3$, and $M_{\rm BH}\pm1\sigma$. $^\ddagger$: $I_{\rm obs}$ exceeds the Schwarzschild maximum $I_{\max}=0.56$ by more than $2\sigma$; the baseline is fitted to the lags only (Section~\ref{sec:illum-limit}).}
\label{tab:fits}
\small
\begin{tabular}{lcccccccc}
\toprule
Source & $\chi^2_{\rm lag}/\nu$ & $\chi^2_I$ & $A$ (s) ($\gamma_c$) & $R_c$ ($r_g$) & $z_c$ ($r_g$) & syst. ($\Delta R_c$, $\Delta z_c$) & flat ($R_c$, $z_c$) & $\langle t\rangle$ ($r_g/c$) \\
\midrule
Mrk 335 (2006)$^\ddagger$ & 1.0/2 & -- & 0 & $9.1^{+1.5}_{-2.0}$ & $3.1^{+2.6}_{-0.7}$ & 1.6, 1.4 & (10.7, 3.9) & 22 \\
Mrk 335 (2009) & 0.2/2 & 0.66 & 262 (2.0) & $11.7^{+1.9}_{-2.6}$ & $6.9^{+2.6}_{-3.1}$ & 3.3, 2.6 & (14.9, 10.8) & 31 \\
Mrk 335 (2015)$^\ddagger$ & 5.9/2 & -- & 0 & $8.1^{+0.9}_{-1.3}$ & $4.1^{+2.1}_{-1.3}$ & 3.5, 3.4 & (7.9, 6.2) & 39 \\
Mrk 335 (2018)$^\ddagger$ & 0.0/2 & -- & 408 (2.0) & $8.6^{+0.5}_{-2.3}$ & $5.0^{+0.5}_{-2.0}$ & 1.1, 3.0 & (9.6, 5.8) & 28 \\
Mrk 335 (2019)$^\ddagger$ & 3.7/2 & -- & 1161 (1.0) & $8.2^{+1.3}_{-0.7}$ & $2.9^{+7.3}_{-0.4}$ & 1.5, 0.8 & (8.2, 2.3) & 21 \\
1H 0707$-$495$^\ddagger$ & 5.0/5 & -- & 12 (2.0) & $8.2^{+2.7}_{-2.4}$ & $3.9^{+0.8}_{-1.0}$ & 3.8, 3.2 & (11.3, 3.8) & 18 \\
IRAS 13224$-$3809$^\ddagger$ & 2.8/3 & -- & 30 (2.0) & $6.3^{+0.7}_{-1.0}$ & $6.0^{+1.5}_{-1.6}$ & 2.6, 1.2 & (8.2, 5.9) & 31 \\
MCG$-$6$-$30$-$15 & 0.7/3 & 0.00 & 15 (0.0) & $9.9^{+1.4}_{-2.1}$ & $6.0^{+2.9}_{-2.1}$ & 5.4, 3.2 & (13.8, 6.0) & 25 \\
\bottomrule
\end{tabular}
\end{table*}

\begin{figure*}
\centering
\begin{tikzpicture}
\begin{groupplot}[group style={group size=4 by 2, horizontal sep=6mm, vertical sep=12mm, xlabels at=edge bottom, ylabels at=edge left, yticklabels at=edge left},
  width=0.235\textwidth, height=0.235\textwidth, xmode=log, ymode=log, enlargelimits=false, axis on top, xtick={3,10,30}, ytick={0.1,1,10}, log ticks with fixed point, colormap name=hamcor, point meta min=0, point meta max=1,
  xlabel={$R\;[r_g]$}, ylabel={$z\;[r_g]$}, title style={font=\scriptsize, yshift=-1mm},
  tick label style={font=\scriptsize}, label style={font=\small}]
\nextgroupplot[title={Mrk 335 (2006)}]
\addplot[matrix plot*, mesh/cols=10, point meta=explicit] table[x=x, y=y, meta=c] {figures/data/mrk335_2006/map_bootmean.dat};
\nextgroupplot[title={Mrk 335 (2009)}]
\addplot[matrix plot*, mesh/cols=10, point meta=explicit] table[x=x, y=y, meta=c] {figures/data/mrk335_2009/map_bootmean.dat};
\nextgroupplot[title={Mrk 335 (2015)}]
\addplot[matrix plot*, mesh/cols=10, point meta=explicit] table[x=x, y=y, meta=c] {figures/data/mrk335_2015/map_bootmean.dat};
\nextgroupplot[title={Mrk 335 (2018)}]
\addplot[matrix plot*, mesh/cols=10, point meta=explicit] table[x=x, y=y, meta=c] {figures/data/mrk335_2018/map_bootmean.dat};
\nextgroupplot[title={Mrk 335 (2019)}]
\addplot[matrix plot*, mesh/cols=10, point meta=explicit] table[x=x, y=y, meta=c] {figures/data/mrk335_2019/map_bootmean.dat};
\nextgroupplot[title={1H 0707$-$495}]
\addplot[matrix plot*, mesh/cols=10, point meta=explicit] table[x=x, y=y, meta=c] {figures/data/1h0707/map_bootmean.dat};
\nextgroupplot[title={IRAS 13224$-$3809}]
\addplot[matrix plot*, mesh/cols=10, point meta=explicit] table[x=x, y=y, meta=c] {figures/data/iras13224/map_bootmean.dat};
\nextgroupplot[title={MCG$-$6$-$30$-$15}, colorbar right, colorbar style={width=2mm, ytick={0,0.5,1}, ylabel={$\sum_\phi w_i/\max$}, ylabel style={font=\scriptsize}, tick label style={font=\scriptsize}}]
\addplot[matrix plot*, mesh/cols=10, point meta=explicit] table[x=x, y=y, meta=c] {figures/data/mcg6/map_bootmean.dat};
\end{groupplot}
\end{tikzpicture}
\caption{Emissivity maps $\sum_\phi w_i(R,z)$ averaged over the 30 bootstrap refits (exact Schwarzschild transfer, $J=1$), normalised to their maximum. With $N_{\rm eff}\simeq3$--$5$ the data constrain the modes of Fig.~\ref{fig:modes} and the centroid; the remaining structure is set by the smoothness prior. The bootstrap scatter is shown in Fig.~\ref{fig:mapscv}. Sources marked $^\ddagger$ in Table~\ref{tab:fits} (all except Mrk~335 2009 and MCG$-$6$-$30$-$15) are fitted to the lags only, because their reflection fraction exceeds the Schwarzschild maximum (Section~\ref{sec:illum-limit}).}
\label{fig:maps}
\end{figure*}

\subsection{Mrk 335 across five epochs}
\label{sec:mrk335}

Between 2006 and 2019 the soft count rate of Mrk~335 dropped by a factor $\sim40$ and the
reflection fraction rose from $\sim2$--$4$ to $\sim7$--$10$ (Table~\ref{tab:spectral}). The band
lag goes from $-1.1\pm0.5\,\rgc$ (2006) to $-9\pm6$ and $-10\pm4\,\rgc$ (2015, 2018) and becomes
positive in 2019, where the continuum lag dominates (Fig.~\ref{fig:mrk335}). The recovered
centroids are $R_c\simeq8$--$12\,\rg$ and $z_c\simeq3$--$7\,\rg$ at all epochs. They are
consistent with each other within the (large) bootstrap intervals, but the data do not have the
sensitivity to detect the contraction of the corona reported for the low states by
\citet{WilkinsGallo2015}; the 2009 solution, the only one constrained
by the illumination fraction, is the most extended.

With four points per epoch and large errors in the low states, this consistency does not
exclude geometric changes. The first version claimed a stable centroid at
$(6.3,0.5)\,\rg$, very close to the disc, and argued against a changing corona; that result
was driven by the over-counted illumination of cells adjacent to the disc
(Section~\ref{sec:illum}) and is not confirmed. The increase of $R_f$ towards low fluxes is,
however, the behaviour expected from a corona that becomes more compact and closer to the
hole \citep{Parker2014,WilkinsGallo2015}, and is the reason why most epochs exceed
$\Imax$.

\subsection{Spin dependence}
\label{sec:spin}

Table~\ref{tab:kerr} repeats the analysis with the Kerr transfer (Section~\ref{sec:kerr}).
At $a=0.998$ the maximum illumination fraction of single cells rises to $0.937$
($R_f\le15$) and that of a lamppost to $0.960$ at $h=1.12\,\rg$ (Fig.~\ref{fig:kerrillum}),
so every observation except IRAS~13224$-$3809 ($\Iobs=0.98$) can be fitted with lags and
illumination together; $\Iobs$ and $\Dil$ are taken from the spectral refit at the same
spin (Table~\ref{tab:spectral}). With the illumination imposed, the coronae of Mrk~335 and
1H~0707$-$495 become compact and close to the hole, $R_c\simeq2.0$--$3.4\,\rg$ and
$z_c\simeq0.6$--$4.7\,\rg$ (Fig.~\ref{fig:kerrmaps}), as expected from the light bending that
reflection fractions of $3$--$20$ require, while MCG$-$6$-$30$-$15, with a moderate
reflection fraction, remains extended ($R_c\simeq16$, $z_c\simeq10\,\rg$). For Mrk~335 and
1H~0707$-$495 the centroid radius is a factor of 3--4 smaller than in Schwarzschild
(Table~\ref{tab:fits}), well beyond the systematic shifts of $1$--$4\,\rg$ quoted there: the
inferred position of the corona, not only its illumination, depends on the spin assumed in
the forward model.

For Mrk~335 in 2015, 2018 and 2019 the compact solution required by the illumination is in
tension with the lags ($\chi^2_{\rm lag}=10.0$, $22.6$ and $11.6$ for 2 dof): the measured soft
lags are longer than a corona at $\sim2\,\rg$ can produce with the measured dilution. Either
the reflection fraction of these low states is overestimated by the spectral model
(Section~\ref{sec:spectra}), or part of the lag in the reverberation band is not
reverberation. The lamppost ($h\simeq1.2$--$3.6\,\rg$ for Mrk~335 and 1H~0707$-$495) remains
preferred by the AIC in all cases except Mrk~335 in 2015 ($\Delta{\rm AIC}=+1.2$, not
significant).

Frame dragging breaks the parity degeneracy of Section~\ref{sec:parity}, but weakly:
reflecting the $a=0.998$ solutions about the observer's meridian plane changes the lags by
$0.2$--$1.6$ per cent of their maximum, comparable to the numerical floor at $a=0$ and far
below the measurement errors. The odd part of the emissivity remains unconstrained by lags
at any spin.

\begin{table*}
\centering
\caption{HAMCOR and lamppost fits with the Kerr transfer on the extended grid. $a=0$: consistency check against the Schwarzschild module ($\chi^2_{\rm lag}$ of Table~\ref{tab:fits} in brackets). $a=0.998$: $I_{\rm obs}$ and dilution from the spectral fit at the same spin; centroids are bootstrap medians and 16--84 percentiles (10 refits). $\Delta{\rm AIC}$ as in Table~\ref{tab:lamppost} (positive favours HAMCOR). $\mathcal{P}$: maximum change of the lags under reflection of the map about the observer's meridian plane, relative to the maximum lag (zero in Schwarzschild; the numerical floor at $a=0$ is $\le1$ per cent). $^\ddagger$: $I_{\rm obs}>I_{\max}$, lags only.}
\label{tab:kerr}
\footnotesize
\setlength{\tabcolsep}{3.5pt}
\begin{tabular}{lccccccccccc}
\toprule
& \multicolumn{2}{c}{$a=0$} & \multicolumn{9}{c}{$a=0.998$} \\ \cmidrule(lr){2-3}\cmidrule(lr){4-12} Source & $\chi^2_{\rm lag}$ (Schw.) & $\Delta$AIC & $I_{\rm obs}$ & $\chi^2_{\rm lag}/\nu$ & $\chi^2_I$ & $R_c$ ($r_g$) & $z_c$ ($r_g$) & $N_{\rm eff}$ & $h_{\rm LP}$ ($r_g$) & $\Delta$AIC & $\mathcal{P}$ (\%) \\
\midrule
Mrk 335 (2006) & 1.03 (1.00) & -3.4 & 0.81 & 1.8/2 & 0.2 & $2.6^{+0.1}_{-0.4}$ & $1.1^{+0.4}_{-0.2}$ & 3.7 & 2.4 & -5.9 & 0.3 \\
Mrk 335 (2009) & 0.14 (0.20) & -4.6 & 0.86 & 0.9/2 & 0.9 & $3.4^{+2.8}_{-1.0}$ & $4.7^{+1.9}_{-1.0}$ & 4.4 & 1.3 & -7.9 & 0.4 \\
Mrk 335 (2015) & 5.72 (5.88) & -3.2 & 0.72 & 10.0/2 & 0.0 & $2.3^{+0.1}_{-0.3}$ & $3.7^{+0.1}_{-0.5}$ & 4.4 & 3.6 & +1.2 & 1.4 \\
Mrk 335 (2018) & 0.03 (0.03) & -5.3 & 0.88 & 22.6/2 & 0.4 & $2.0^{+0.1}_{-0.0}$ & $1.0^{+0.1}_{-0.1}$ & 4.0 & 1.6 & -17.1 & 0.5 \\
Mrk 335 (2019) & 3.73 (3.70) & -7.5 & 0.95 & 11.6/2 & 5.2 & $2.2^{+0.1}_{-0.3}$ & $0.9^{+0.9}_{-0.1}$ & 3.8 & 1.2 & -18.6 & 1.6 \\
1H 0707$-$495 & 5.05 (5.03) & -4.8 & 0.93 & 5.3/5 & 3.5 & $2.2^{+0.0}_{-0.0}$ & $0.6^{+0.1}_{-0.1}$ & 4.5 & 1.4 & -8.1 & 0.2 \\
IRAS 13224$-$3809$^\ddagger$ & 2.92 (2.82) & -3.3 & 0.98 & 3.0/3 & -- & $8.7^{+1.8}_{-0.8}$ & $10.1^{+1.5}_{-1.3}$ & 3.6 & 2.7 & -2.6 & 0.4 \\
MCG$-$6$-$30$-$15 & 0.66 (0.67) & -4.8 & 0.45 & 0.6/3 & 1.6 & $15.9^{+2.1}_{-5.4}$ & $10.2^{+2.9}_{-1.3}$ & 3.9 & 14.6 & -2.8 & 0.0 \\
\bottomrule
\end{tabular}
\end{table*}

\begin{figure}
\centering
\begin{tikzpicture}
\begin{axis}[width=0.95\columnwidth, height=0.62\columnwidth, xmode=log, xmin=1.1, xmax=40,
  ymin=0.2, ymax=1.3, ytick={0.2,0.4,0.6,0.8,1}, xtick={2,5,10,20,40}, log ticks with fixed point,
  xlabel={$h\;[r_g]$}, ylabel={$I$}, legend style={font=\scriptsize, at={(0.98,0.98)}, anchor=north east},
  tick label style={font=\scriptsize}, label style={font=\small}]
\fill[orange!25] (axis cs:1.1,0.72) rectangle (axis cs:40,0.98);
\addplot[thick, black] table[x=h, y=I] {figures/data/kerr/lamppost_I_a0.dat};
\addlegendentry{lamppost, $a=0$}
\addplot[thick, blue!70!black] table[x=h, y=I] {figures/data/kerr/lamppost_I_a0998.dat};
\addlegendentry{lamppost, $a=0.998$}
\addplot[dashed, black, domain=1.1:40] {0.561};
\addlegendentry{$I_{\max}$ cells, $a=0$}
\addplot[dashed, blue!70!black, domain=1.1:40] {0.937};
\addlegendentry{$I_{\max}$ cells, $a=0.998$}
\end{axis}
\end{tikzpicture}
\caption{Illumination fraction of an on-axis point source (inclination $30^\circ$, disc from the ISCO to $100\,r_g$) for $a=0$ and $a=0.998$ (Kerr ray tracing), and the maximum over single cells of the extended grid (dashed). The shaded band spans the illumination fractions measured at $a=0.998$ for Mrk~335, 1H~0707$-$495 and IRAS~13224$-$3809 ($0.72$--$0.98$).}
\label{fig:kerrillum}
\end{figure}

\subsection{Energy-resolved reverberation}
\label{sec:eres-results}

Table~\ref{tab:energy} reports the lag-energy spectra predicted by, and fitted with, the
energy-resolved transfer (Section~\ref{sec:eres}). The three low-flux epochs of Mrk~335
(2015--2019) have median raw coherence $0.02$--$0.03$ in the lag-energy channels, and even a
constant lag gives $\chi^2\simeq200$--$260$ for 10 dof: at such low coherence and with four
or five segments the Gaussian lag errors underestimate the scatter
\citep{Epitropakis2016}, and we do not fit these spectra. For the other five observations
we drop the few channels with raw coherence below $0.01$.

The energy shifts change the predicted lag-energy spectra of the lag-frequency solutions
little at $a=0$, where the disc ends at $6\,\rg$ (Table~\ref{tab:energy}), and by up to
$\sim20$ per cent in $\chi^2$ at $a=0.998$, where the innermost, most redshifted disc responds
first in the red wing of the iron line. Fitting the lag-frequency and lag-energy spectra
jointly, HAMCOR reaches a lower $\chi^2$ than the lamppost in every case, by $3$ to $24$, and
the number of constrained parameters rises from $3$--$5$ to $12$--$16$: the lag-energy
spectrum carries geometric information that the lag-frequency spectrum does not. With
current data the gain does not pay for the additional parameters: the AIC prefers the
lamppost by $9$--$20$ at $a=0$ and by $12$--$20$ at $a=0.998$ for four of the five
observations, and is neutral for MCG$-$6$-$30$-$15 at $a=0.998$ ($+0.7$). The
energy-resolved comparison therefore applies to five of the eight observations; for the
three low-flux epochs of Mrk~335 the lag-frequency comparison (Tables~\ref{tab:lamppost}
and~\ref{tab:kerr}) remains the only one.

For MCG$-$6$-$30$-$15 the two models are indistinguishable. Its illumination fraction is
moderate ($\Iobs=0.45$ at $a=0.998$), so the illumination does not push either model towards
the hole: the lamppost settles at $h\simeq9$--$15\,\rg$ and HAMCOR at $R_c\simeq16$,
$z_c\simeq10\,\rg$, and both reproduce a lag-energy spectrum that is consistent with a
constant ($\chi^2=20$ for 14 dof). The discriminating power found for Mrk~335 in 2006
arises only when a large, precisely measured reflection fraction fixes the lamppost
height and the lag-energy spectrum is flat.

The exception is Mrk~335 in 2006 at $a=0.998$ ($\Delta{\rm AIC}=+27.8$; Fig.~\ref{fig:eres}),
and it is marginal. The measured spectrum is nearly flat ($\chi^2=21.5$ for a constant). A
lamppost at the height that matches the illumination fraction, $\Iobs=0.81\pm0.02$
($h\simeq2.4\,\rg$), predicts a strong soft excess and iron-line structure
($\chi^2_{\rm le}=88.5$ for 15 points); a lamppost just outside the horizon ($h\simeq1.2\,\rg$)
reproduces the flat spectrum ($\chi^2_{\rm le}=21.0$) but over-predicts the illumination
($I=0.95$, $\chi^2_I=46$). An extended HAMCOR corona satisfies both ($\chi^2_{\rm le}=14.3$,
$\chi^2_I=0.6$). The preference therefore rests entirely on the precision of $\Iobs$:
inflating $\sigma_I=0.021$ by factors of 1.5, 2 and 3 gives $\Delta{\rm AIC}=+2.6$, $-6.2$ and
$-12.6$. To measure the model dependence of $\Iobs$ we refitted the spectrum at $a=0.998$
with six alternative models (Table~\ref{tab:robust}): a fixed emissivity break, no distant
reflector, no additional Gaussian, \textsc{relxillCp} with free density, and inclinations of
$30^\circ$ and $60^\circ$. They give $I=0.76$--$0.95$, a half-spread of $0.095$, three times
the precision the result requires; the statistical error of the baseline refit alone is
comparable. With $\sigma_I\gtrsim0.06$ the lamppost is preferred ($\Delta{\rm AIC}\lesssim-12$).
The present data therefore do not exclude a lamppost for this observation either; the case
identifies the combination of observables --- a reflection fraction measured to a few per
cent independently of the spectral model, together with a flat, high-coherence lag-energy
spectrum --- that could do so.

\begin{table}
\centering
\caption{Model dependence of the illumination fraction of Mrk~335 (2006) at $a=0.998$. Each
row refits the spectrum with one change to the baseline model
(\textsc{TBabs*(relxill+xillver+zgauss)}, $i=45^\circ$, broken emissivity with free $R_{\rm br}$).
$\sigma_I$ is the larger of the covariance-draw and profile errors.}
\label{tab:robust}
\small
\setlength{\tabcolsep}{4pt}
\begin{tabular}{lcccc}
\toprule
Variant & $\chi^2/\nu$ & $R_f$ & $I$ & $\sigma_I$ \\
\midrule
baseline & 220/168 & 4.44 & 0.816 & 0.188 \\
$R_{\rm br}=6\,\rg$ fixed & 259/169 & 3.19 & 0.762 & 0.047 \\
no distant reflector & 359/169 & 20.0$^a$ & 0.952 & 0.470 \\
no Gaussian & 240/170 & 8.19 & 0.891 & 0.322 \\
\textsc{relxillCp}, free $n_{\rm e}$ & 219/167 & 7.56 & 0.883 & 0.347 \\
$i=30^\circ$ & 197/168 & 5.66 & 0.850 & 0.047 \\
$i=60^\circ$ & 402/168 & 5.70 & 0.851 & 0.005 \\
\bottomrule
\multicolumn{5}{l}{\footnotesize $^a$ at the upper bound of the fit.}
\end{tabular}
\end{table}

\begin{table*}
\centering
\caption{Energy-resolved reverberation. $\bar\gamma^2$: median raw coherence of the lag-energy channels. PPC: $\chi^2$ of the lag-energy spectrum predicted by the lag-frequency fits with the energy-resolved transfer (HAMCOR/lamppost/zero lag, each with a free offset; $a=0$). Joint fits: $\chi^2_{\rm lag}+\chi^2_{\rm le}$ for HAMCOR (H) and the lamppost (LP; the illumination term, used at $a=0.998$ except for IRAS~13224$-$3809, is not shown), $N_{\rm eff}$ from the lag-frequency data alone and from the joint data, and $\Delta{\rm AIC}$ (positive favours HAMCOR). Observations with $\bar\gamma^2<0.05$ have unreliable lag-energy errors and are not fitted.}
\label{tab:energy}
\footnotesize
\setlength{\tabcolsep}{3.5pt}
\begin{tabular}{lcccccccccc}
\toprule
& & & \multicolumn{4}{c}{joint fit, $a=0$} & \multicolumn{4}{c}{joint fit, $a=0.998$} \\ \cmidrule(lr){4-7}\cmidrule(lr){8-11} Source & $\bar\gamma^2$ & PPC H/LP/0 (dof) & H & LP & $N_{\rm eff}$ & $\Delta$AIC & H & LP & $N_{\rm eff}$ & $\Delta$AIC \\
\midrule
Mrk 335 (2006) & 0.32 & 40/116/21 (15) & 2.5+10.8 & 2.7+18.0 & 3.1$\to$14.7 & -20.0 & 2.4+14.3 & 3.1+21.0 & 3.3$\to$13.5 & +27.8 \\
Mrk 335 (2009) & 0.08 & 78/36/34 (14) & 2.8+30.7 & 1.2+35.0 & 3.8$\to$12.8 & -19.9 & 2.7+31.9 & 2.8+34.1 & 3.6$\to$11.7 & -19.9 \\
Mrk 335 (2015) & 0.02 & 255/274/264 (10) & -- & -- & -- & -- & -- & -- & -- & -- \\
Mrk 335 (2018) & 0.02 & 212/205/238 (10) & -- & -- & -- & -- & -- & -- & -- & -- \\
Mrk 335 (2019) & 0.03 & 240/220/205 (10) & -- & -- & -- & -- & -- & -- & -- & -- \\
1H 0707$-$495 & 0.13 & 47/88/40 (13) & 5.7+28.0 & 17.7+37.7 & 4.6$\to$16.5 & -9.2 & 5.8+27.0 & 6.7+42.1 & 4.5$\to$16.2 & -17.6 \\
IRAS 13224$-$3809 & 0.11 & 45/65/25 (13) & 3.5+17.5 & 7.0+24.0 & 3.8$\to$14.3 & -16.5 & 3.6+19.5 & 6.7+28.2 & 3.5$\to$12.9 & -12.2 \\
MCG$-$6$-$30$-$15 & 0.18 & 20/21/20 (14) & 0.8+11.5 & 0.9+19.7 & 3.1$\to$13.7 & -16.8 & 0.7+12.1 & 1.1+27.0 & 3.1$\to$13.0 & +0.7 \\
\bottomrule
\end{tabular}
\end{table*}

\begin{figure*}
\centering
\begin{tikzpicture}
\begin{groupplot}[group style={group size=5 by 2, horizontal sep=9mm, vertical sep=12mm, xlabels at=edge bottom},
  width=0.23\textwidth, height=0.2\textwidth, xmode=log, xmin=0.3, xmax=10,
  xtick={0.5,1,2,5}, log ticks with fixed point, xlabel={$E$ [keV]},
  tick label style={font=\scriptsize}, label style={font=\small}, unbounded coords=jump,
  title style={font=\scriptsize, yshift=-1mm}]
\nextgroupplot[title={Mrk 335 (2006)}, ylabel={lag [s], $a=0$}]
\addplot[only marks, mark=*, mark size=1pt, black, error bars/.cd, y dir=both, y explicit] table[x=E, y=data, y error=sig] {figures/data/mrk335_2006/le_eres_schwarzschild.dat};
\addplot[thick, blue!70!black] table[x=E, y=hamcor_joint] {figures/data/mrk335_2006/le_eres_schwarzschild.dat};
\addplot[thick, dashed, red!70!black] table[x=E, y=lamppost_joint] {figures/data/mrk335_2006/le_eres_schwarzschild.dat};
\nextgroupplot[title={Mrk 335 (2009)}]
\addplot[only marks, mark=*, mark size=1pt, black, error bars/.cd, y dir=both, y explicit] table[x=E, y=data, y error=sig] {figures/data/mrk335_2009/le_eres_schwarzschild.dat};
\addplot[thick, blue!70!black] table[x=E, y=hamcor_joint] {figures/data/mrk335_2009/le_eres_schwarzschild.dat};
\addplot[thick, dashed, red!70!black] table[x=E, y=lamppost_joint] {figures/data/mrk335_2009/le_eres_schwarzschild.dat};
\nextgroupplot[title={1H 0707$-$495}]
\addplot[only marks, mark=*, mark size=1pt, black, error bars/.cd, y dir=both, y explicit] table[x=E, y=data, y error=sig] {figures/data/1h0707/le_eres_schwarzschild.dat};
\addplot[thick, blue!70!black] table[x=E, y=hamcor_joint] {figures/data/1h0707/le_eres_schwarzschild.dat};
\addplot[thick, dashed, red!70!black] table[x=E, y=lamppost_joint] {figures/data/1h0707/le_eres_schwarzschild.dat};
\nextgroupplot[title={IRAS 13224$-$3809}]
\addplot[only marks, mark=*, mark size=1pt, black, error bars/.cd, y dir=both, y explicit] table[x=E, y=data, y error=sig] {figures/data/iras13224/le_eres_schwarzschild.dat};
\addplot[thick, blue!70!black] table[x=E, y=hamcor_joint] {figures/data/iras13224/le_eres_schwarzschild.dat};
\addplot[thick, dashed, red!70!black] table[x=E, y=lamppost_joint] {figures/data/iras13224/le_eres_schwarzschild.dat};
\nextgroupplot[title={MCG$-$6$-$30$-$15}]
\addplot[only marks, mark=*, mark size=1pt, black, error bars/.cd, y dir=both, y explicit] table[x=E, y=data, y error=sig] {figures/data/mcg6/le_eres_schwarzschild.dat};
\addplot[thick, blue!70!black] table[x=E, y=hamcor_joint] {figures/data/mcg6/le_eres_schwarzschild.dat};
\addplot[thick, dashed, red!70!black] table[x=E, y=lamppost_joint] {figures/data/mcg6/le_eres_schwarzschild.dat};
\nextgroupplot[title={}, ylabel={lag [s], $a=0.998$}]
\addplot[only marks, mark=*, mark size=1pt, black, error bars/.cd, y dir=both, y explicit] table[x=E, y=data, y error=sig] {figures/data/mrk335_2006/le_eres_kerr.dat};
\addplot[thick, blue!70!black] table[x=E, y=hamcor_joint] {figures/data/mrk335_2006/le_eres_kerr.dat};
\addplot[thick, dashed, red!70!black] table[x=E, y=lamppost_joint] {figures/data/mrk335_2006/le_eres_kerr.dat};
\nextgroupplot[title={}]
\addplot[only marks, mark=*, mark size=1pt, black, error bars/.cd, y dir=both, y explicit] table[x=E, y=data, y error=sig] {figures/data/mrk335_2009/le_eres_kerr.dat};
\addplot[thick, blue!70!black] table[x=E, y=hamcor_joint] {figures/data/mrk335_2009/le_eres_kerr.dat};
\addplot[thick, dashed, red!70!black] table[x=E, y=lamppost_joint] {figures/data/mrk335_2009/le_eres_kerr.dat};
\nextgroupplot[title={}]
\addplot[only marks, mark=*, mark size=1pt, black, error bars/.cd, y dir=both, y explicit] table[x=E, y=data, y error=sig] {figures/data/1h0707/le_eres_kerr.dat};
\addplot[thick, blue!70!black] table[x=E, y=hamcor_joint] {figures/data/1h0707/le_eres_kerr.dat};
\addplot[thick, dashed, red!70!black] table[x=E, y=lamppost_joint] {figures/data/1h0707/le_eres_kerr.dat};
\nextgroupplot[title={}]
\addplot[only marks, mark=*, mark size=1pt, black, error bars/.cd, y dir=both, y explicit] table[x=E, y=data, y error=sig] {figures/data/iras13224/le_eres_kerr.dat};
\addplot[thick, blue!70!black] table[x=E, y=hamcor_joint] {figures/data/iras13224/le_eres_kerr.dat};
\addplot[thick, dashed, red!70!black] table[x=E, y=lamppost_joint] {figures/data/iras13224/le_eres_kerr.dat};
\nextgroupplot[title={}]
\addplot[only marks, mark=*, mark size=1pt, black, error bars/.cd, y dir=both, y explicit] table[x=E, y=data, y error=sig] {figures/data/mcg6/le_eres_kerr.dat};
\addplot[thick, blue!70!black] table[x=E, y=hamcor_joint] {figures/data/mcg6/le_eres_kerr.dat};
\addplot[thick, dashed, red!70!black] table[x=E, y=lamppost_joint] {figures/data/mcg6/le_eres_kerr.dat};
\end{groupplot}
\end{tikzpicture}
\caption{Lag-energy spectra in the reverberation band with the joint (lag-frequency + lag-energy) fits using the energy-resolved transfer: HAMCOR (solid) and lamppost (dashed); top $a=0$ (Schwarzschild), bottom $a=0.998$ (Kerr, illumination fraction imposed except for IRAS~13224$-$3809). Channels with raw coherence below $0.01$ are omitted. The three low-flux epochs of Mrk~335 are not shown (median coherence below $0.05$; Section~\ref{sec:eres-results}).}
\label{fig:eres}
\end{figure*}

\section{Discussion}
\label{sec:discussion}

\subsection{Comparison with a lamppost}
\label{sec:lamppost}

We fit an on-axis point source to the same data, with the same estimator, dilution,
continuum lag, exact transfer and treatment of the illumination
(Table~\ref{tab:lamppost}, Fig.~\ref{fig:lamppost}). The lamppost describes the lags as
well as HAMCOR, with $\chi^2$ larger by at most 2.3; heights are low ($2.6$--$7\,\rg$) for six
observations, with broad or bimodal $\chi^2(h)$ curves. Penalising the $N_{\rm eff}\simeq3$--$5$
effective parameters of HAMCOR against one for the lamppost, the Akaike criterion
\citep{Akaike1974} prefers the lamppost in all eight cases ($\Delta{\rm AIC}=-3.5$ to $-7.0$).
With the Kerr transfer at $a=0.998$ it is preferred in seven of eight, the exception being
Mrk~335 in 2015 ($+1.2$, not significant; Section~\ref{sec:spin}); when the lag-energy spectra
are included (Section~\ref{sec:eres-results}) it remains preferred for four of the five usable
observations at $a=0$ and $a=0.998$, except, marginally, Mrk~335 in 2006 at $a=0.998$, while for
MCG$-$6$-$30$-$15 at $a=0.998$ the two models are indistinguishable ($+0.7$).
In an earlier run, without the emissivity-model systematic on $\Dil$
(Section~\ref{sec:spectra}), HAMCOR was preferred for Mrk~335 in 2015; that preference rested
on a formally poor fit ($\chi^2_{\rm lag}=13.5$ for 2 dof) and disappears once the dilution
uncertainty is propagated, so we do not regard it as evidence for an extended corona. The
present lag-frequency data are therefore consistent with an extended corona but do not require it. An extended solution
and a low lamppost produce similar first moments of the response: the low-frequency lag
measures $\Dil\langle t\rangle$ (equation~\ref{eq:lowf}), and the frequency dependence
over less than one decade carries little additional information.

The lag-energy spectra in the reverberation band are not fitted and provide an independent
test (Appendix~\ref{app:lagenergy}). The HAMCOR predictions are better than the lamppost ones for Mrk~335 in 2006 and 2015,
the two NLS1s and MCG$-$6$-$30$-$15, but none of the models is better than a constant lag for
most observations (PPC$_0$ in Table~\ref{tab:lamppost}); for Mrk~335 in 2006 the HAMCOR
prediction ($\chi^2=41$ for 15 dof) is worse than zero lag (21). Adding the energy shifts of
the reflected spectrum (Section~\ref{sec:eres}) changes these predictions little; fitting
the lag-energy spectra jointly with the lag-frequency spectra
(Section~\ref{sec:eres-results}) raises $N_{\rm eff}$ to $12$--$16$; at maximal spin it
would disfavour the lamppost for Mrk~335 in 2006 only if the reflection fraction were known
to better than $0.03$, three times better than its model dependence
(Table~\ref{tab:robust}). This comparison covers five of the eight
observations; the three low-flux epochs of Mrk~335 enter only through their lag-frequency
spectra.

\begin{table*}
\centering
\caption{Lamppost (LP) and HAMCOR (H) fitted to identical data with the same estimator, dilution, continuum lag and spacetime. $N_{\rm eff}$: number of data-constrained modes of the HAMCOR solution (Fisher spectrum). $\Delta{\rm AIC}={\rm AIC}_{\rm LP}-{\rm AIC}_{\rm H}$ (positive favours HAMCOR). PPC: $\chi^2$/dof of the lag-energy spectrum (not fitted) for the HAMCOR and lamppost predictions and for zero lag, each with a free additive offset.}
\label{tab:lamppost}
\small
\begin{tabular}{lcccccccc}
\toprule
Source & $h_{\rm LP}$ ($r_g$) & $\chi^2_{\rm LP}$ & $\chi^2_{\rm H}$ & $N_{\rm eff}$ & $\Delta$AIC & PPC$_{\rm H}$ & PPC$_{\rm LP}$ & PPC$_0$ \\
\midrule
Mrk 335 (2006) & $2.6^{+37.4}_{-0.1}$ & 2.0 & 1.0 & 3.4 & -3.9 & 41/15 & 111/15 & 21/15 \\
Mrk 335 (2009) & $31.0^{+3.0}_{-26.3}$ & 2.6 & 0.9 & 4.8 & -5.8 & 77/15 & 36/15 & 34/15 \\
Mrk 335 (2015) & $12.8^{+0.0}_{-0.9}$ & 8.2 & 5.9 & 3.9 & -3.5 & 265/15 & 289/15 & 270/15 \\
Mrk 335 (2018) & $4.6^{+0.1}_{-0.6}$ & 0.2 & 0.0 & 3.8 & -5.4 & 213/15 & 211/15 & 241/15 \\
Mrk 335 (2019) & $7.1^{+0.3}_{-0.0}$ & 1.9 & 3.7 & 3.6 & -7.0 & 244/15 & 230/15 & 209/15 \\
1H 0707$-$495 & $2.6^{+3.0}_{-0.1}$ & 5.9 & 5.0 & 4.1 & -5.3 & 50/15 & 78/15 & 45/15 \\
IRAS 13224$-$3809 & $11.9^{+0.3}_{-9.4}$ & 4.4 & 2.8 & 3.8 & -4.1 & 40/15 & 62/15 & 29/15 \\
MCG$-$6$-$30$-$15 & $5.4^{+1.4}_{-0.9}$ & 0.9 & 0.7 & 3.4 & -4.6 & 17/15 & 20/15 & 20/15 \\
\bottomrule
\end{tabular}
\end{table*}

\begin{figure*}
\centering
\begin{tikzpicture}
\begin{groupplot}[group style={group size=4 by 2, horizontal sep=10mm, vertical sep=13mm, xlabels at=edge bottom},
  width=0.27\textwidth, height=0.22\textwidth, xmode=log, xmin=2.5, xmax=40,
  xtick={3,10,30}, log ticks with fixed point, xlabel={$h$ [$r_g$]},
  tick label style={font=\scriptsize}, label style={font=\small},
  title style={font=\scriptsize, yshift=-1mm}]
\nextgroupplot[title={Mrk 335 (2006)}, ylabel={$\chi^2_{\rm lag}$}]
\addplot[thick, black] table[x=h, y=chi2_lag] {figures/data/mrk335_2006/lamppost_chi2_schwarzschild.dat};
\addplot[thick, dashed, gray] table[x=h, y=chi2_lag] {figures/data/mrk335_2006/lamppost_chi2_flat.dat};
\nextgroupplot[title={Mrk 335 (2009)}]
\addplot[thick, black] table[x=h, y=chi2_lag] {figures/data/mrk335_2009/lamppost_chi2_schwarzschild.dat};
\addplot[thick, dashed, gray] table[x=h, y=chi2_lag] {figures/data/mrk335_2009/lamppost_chi2_flat.dat};
\nextgroupplot[title={Mrk 335 (2015)}]
\addplot[thick, black] table[x=h, y=chi2_lag] {figures/data/mrk335_2015/lamppost_chi2_schwarzschild.dat};
\addplot[thick, dashed, gray] table[x=h, y=chi2_lag] {figures/data/mrk335_2015/lamppost_chi2_flat.dat};
\nextgroupplot[title={Mrk 335 (2018)}]
\addplot[thick, black] table[x=h, y=chi2_lag] {figures/data/mrk335_2018/lamppost_chi2_schwarzschild.dat};
\addplot[thick, dashed, gray] table[x=h, y=chi2_lag] {figures/data/mrk335_2018/lamppost_chi2_flat.dat};
\nextgroupplot[title={Mrk 335 (2019)}, ylabel={$\chi^2_{\rm lag}$}]
\addplot[thick, black] table[x=h, y=chi2_lag] {figures/data/mrk335_2019/lamppost_chi2_schwarzschild.dat};
\addplot[thick, dashed, gray] table[x=h, y=chi2_lag] {figures/data/mrk335_2019/lamppost_chi2_flat.dat};
\nextgroupplot[title={1H 0707$-$495}]
\addplot[thick, black] table[x=h, y=chi2_lag] {figures/data/1h0707/lamppost_chi2_schwarzschild.dat};
\addplot[thick, dashed, gray] table[x=h, y=chi2_lag] {figures/data/1h0707/lamppost_chi2_flat.dat};
\nextgroupplot[title={IRAS 13224$-$3809}]
\addplot[thick, black] table[x=h, y=chi2_lag] {figures/data/iras13224/lamppost_chi2_schwarzschild.dat};
\addplot[thick, dashed, gray] table[x=h, y=chi2_lag] {figures/data/iras13224/lamppost_chi2_flat.dat};
\nextgroupplot[title={MCG$-$6$-$30$-$15}]
\addplot[thick, black] table[x=h, y=chi2_lag] {figures/data/mcg6/lamppost_chi2_schwarzschild.dat};
\addplot[thick, dashed, gray] table[x=h, y=chi2_lag] {figures/data/mcg6/lamppost_chi2_flat.dat};
\end{groupplot}
\end{tikzpicture}
\caption{Lamppost: $\chi^2$ of the lag-frequency data as a function of source height, with the same estimator, dilution and profiled continuum lag as HAMCOR. Solid: exact Schwarzschild transfer; dashed: flat space.}
\label{fig:lamppost}
\end{figure*}

\subsection{Comparison with the literature for Mrk 335}
\label{sec:literature}

Table~\ref{tab:literature} compares our results for Mrk~335 with previous work. Our soft
lag in 2006, $-73\pm32$\,s averaged over 2--7$\times10^{-4}$\,Hz, is consistent with the
$-155\pm57$\,s at $\sim2\times10^{-4}$\,Hz of \citet{Kara2013}. Lamppost fits of spectra and
lags give low heights, $h\simeq3\,\rg$ \citep{Chainakun2016} and $h<2.3$--$5\,\rg$ with
NuSTAR \citep{Parker2014}, in agreement with our lamppost fits. Emissivity-profile analyses
find a corona extending over the inner disc in the high state, out to $26^{+10}_{-7}\,\rg$ in
2006, contracting to $\lesssim12\,\rg$ in 2009 and $\lesssim5\,\rg$ in the low state
\citep{WilkinsGallo2015}; our centroids at $R_c\simeq9$--$12\,\rg$ are compatible with the
high and intermediate states, while our data do not have the sensitivity to detect the
contraction in the low states.

\subsection{Limitations}
\label{sec:limitations}

\textit{Spin.} The baseline is Schwarzschild, and the Kerr analysis
(Section~\ref{sec:spin}) is limited to $a=0$ and $0.998$ with static emitters; intermediate
spins, a spin fitted jointly with the spectra, and orbiting or outflowing coronal material
are not explored.

\textit{Spectral information.} The energy-resolved transfer uses one rest-frame
reflection spectrum per observation (single ionisation and emission angle), and neglects
the energy shift of the illuminating photons and radial ionisation gradients, which
full reverberation codes include \citep{Ingram2019,Mastroserio2018}. The lag-energy
spectra of the low states of Mrk~335 are not usable with Gaussian errors; simulated error
distributions would allow including them.

\textit{Illumination--lag tension.} In three epochs of Mrk~335 the illumination at
$a=0.998$ requires a compact corona while the lags require longer delays
(Section~\ref{sec:spin}). We have not resolved this tension; it may indicate that the
reflection fractions of the low states are overestimated by the spectral model, or that
part of the lag in the reverberation band is not reverberation.

\textit{Isotropic static emission.} Cells emit isotropically in their static frame.
Outflowing or anisotropic coronae change both the illumination and the delays and could
account for part of the large reflection fractions.

\textit{Spectral models.} The 2018--2019 states of Mrk~335 and the two NLS1s are not fitted
satisfactorily by any model tried here. $\Dil$ is robust to the model choices we tested, but
the reflection fractions of these observations should be regarded as model-dependent.

\textit{Information content.} With $N_{\rm eff}\simeq3$--$5$ from the lag-frequency
spectra ($12$--$16$ with the lag-energy spectra) the method returns robust
centroids and mean delays but not detailed maps; this is a property of the data, shared by
any inversion of the same lags, and should be kept in mind when interpreting
Fig.~\ref{fig:maps}.

\section{Conclusions}
\label{sec:conclusions}

\begin{enumerate}
\item HAMCOR infers a non-parametric coronal emissivity from reverberation lags by
minimising a Hamiltonian that combines a smoothness (Gaussian Markov random field) prior,
a lag-frequency likelihood evaluated through the same estimator as the data, and the
measured illumination fraction. Photon transfer is computed by exact geodesic
integration in Schwarzschild spacetime; dilution by the continuum and the intrinsic
continuum lag are part of the model.
\item Lag data in one decade of frequency constrain about four combinations of the
emissivity. Centroids are recovered within $2$--$6\,\rg$; maps only partially. Fitting
Schwarzschild data in flat space biases the centroid by $5$--$9\,\rg$, although the mean delay
changes by only 12--14 per cent.
\item No isotropic corona around a non-spinning hole with the disc at the ISCO can produce
$R_f>1.3$. Six of eight observations exceed this limit, a timing analogue of the argument
for rapid spin from reflection fractions. At $a=0.998$ seven become reachable, and the
illumination then requires compact coronae ($R_c\simeq2$--$3\,\rg$) for Mrk~335 and
1H~0707$-$495, in tension with the lags in the three low-flux epochs of Mrk~335.
\item For Mrk~335 (five epochs), 1H~0707$-$495, IRAS~13224$-$3809 and MCG$-$6$-$30$-$15 the lags
are reproduced by extended coronae with centroids at $R_c\simeq6$--$12\,\rg$,
$z_c\simeq3$--$7\,\rg$. A low lamppost fits equally well and is preferred by the Akaike
criterion; the data are consistent with, but do not require, an extended corona.
\item With the energy shifts of the reflected spectrum included, the lag-energy spectrum
becomes a constraint and raises the number of constrained parameters from $\sim4$ to
$12$--$16$. A lamppost allowed down to the horizon remains preferred; the only exception,
Mrk~335 in 2006 at maximal spin ($\Delta{\rm AIC}=+28$), requires $\sigma_I\lesssim0.03$, while
alternative spectral models spread its illumination fraction by $\pm0.1$; it is therefore
not robust. The energy-resolved comparison is possible for
five of the eight observations. Excluding a lamppost will require observations that combine
a large reflection fraction, measured to a few per cent with independent spectral models,
with a high-coherence lag-energy spectrum.
\item The odd part of the emissivity under reflection about the observer's meridian plane
is invisible to lags and reflection fractions in Schwarzschild spacetime; frame dragging
at $a=0.998$ breaks it only at the per cent level, so X-ray polarimetry
\citep{Krawczynski2022} remains the natural observable.
\end{enumerate}

\begin{acknowledgements}

This work is based on observations obtained with \xmm{}, an ESA science mission with
instruments and contributions directly funded by ESA Member States and NASA. This research
has made use of data and software provided by the High Energy Astrophysics Science Archive
Research Center (HEASARC). We acknowledge the use of \textsc{sas} 22.1.0
\citep{Gabriel2004}, \textsc{heasoft} 6.36 and \textsc{xspec} 12.15.1
\citep{HEASoft2014,Arnaud1996}, \textsc{relxill} 2.9.0 \citep{Garcia2014,Dauser2016},
\textsc{NumPy} \citep{Harris2020}, \textsc{SciPy} \citep{Virtanen2020} and \textsc{Astropy}
\citep{Astropy2022}. Figures were produced with \textsc{pgfplots}.

The \xmm{} data are public in the \xmm{} Science Archive (\url{https://nxsa.esac.esa.int})
under the observation identifiers of Table~\ref{tab:obs}. The HAMCOR code, including the
reduction, lag, spectral-fitting and figure scripts that produce every number and figure
of this paper, is available at \url{https://github.com/fbuffoli95/HAMCOR}.
\end{acknowledgements}

\bibliographystyle{aa}
\bibliography{references}

\begin{appendix}

\section{Nomenclature}
\label{app:nomenclature}
\begin{table*}
\centering
\caption{Symbols used in this paper, with units and the section where they are defined.}
\label{tab:nomenclature}
\small
\begin{tabular}{@{}p{0.17\textwidth}p{0.56\textwidth}p{0.10\textwidth}p{0.08\textwidth}@{}}
\toprule
Symbol & Definition & Units & Section \\
\midrule
\multicolumn{4}{@{}l}{\textit{Scales and geometry}}\\
$\Mbh$ & black-hole mass & M$_\odot$ & \ref{sec:obs} \\
$\rg$, $t_g$ & $G\Mbh/c^2$, $G\Mbh/c^3=4.93\times10^{-6}(\Mbh/{\rm M}_\odot)$\,s & cm, s & \ref{sec:intro} \\
$\theta_{\rm obs}$, $\nhat$ & inclination; unit vector to the observer $(\sin\theta_{\rm obs},0,\cos\theta_{\rm obs})$ & deg, -- & \ref{sec:repr} \\
$(R_i,z_i,\phi_i)$, $\bm x_i$, $r_i$ & cylindrical and Cartesian position, radius of coronal cell $i$ & $\rg$ & \ref{sec:repr} \\
$N$; $n_R,n_z,n_\phi$ & number of cells ($800$); cells per coordinate ($10,10,8$) & -- & \ref{sec:repr} \\
$\bm x_k$, $r_{\rm in}$ & position of disc element $k$; inner disc radius ($6\,\rg$) & $\rg$ & \ref{sec:repr} \\
$w_i$, $\bm w$, $\Delta^{N-1}$ & emissivity weight; emissivity vector; probability simplex & -- & \ref{sec:repr} \\
$(R_c,z_c)$ & emissivity centroid $\sum_iw_i(R_i,z_i)$ & $\rg$ & \ref{sec:optim} \\
\multicolumn{4}{@{}l}{\textit{Photon transfer}}\\
$u$, $b$, $h$, $\lambda$ & $1/r$; impact parameter; angular momentum and parameter of the orbit equation (\ref{eq:geodesic}) & -- & \ref{sec:gr} \\
$\Omega_{ik}$; $a_i$ & solid angle of the photons from cell $i$ reaching disc element $k$; $\sum_k\Omega_{ik}$ & sr & \ref{sec:transfer} \\
$E_i$ & solid angle of the photons from cell $i$ escaping to infinity & sr & \ref{sec:gr} \\
$t_{\rm SP}$, $T(\bm x)$ & coordinate time from source to disc; arrival-time offset to the observer (eq.~\ref{eq:T}) & $\rgc$ & \ref{sec:gr} \\
$L_{ik}$ & delay of reflected relative to direct photons (eq.~\ref{eq:lagmatrix}) & $\rgc$ & \ref{sec:gr} \\
$S(A,B)$, $S_\infty$ & weak-field Shapiro delay of a segment and of a leg to infinity & $\rgc$ & \ref{app:weak} \\
\multicolumn{4}{@{}l}{\textit{Response and observables}}\\
$\psi(t)$, $\tilde\psi(\nu)$ & impulse response (eq.~\ref{eq:psi}) and transfer function (eq.~\ref{eq:psitilde}) & $t_g^{-1}$, -- & \ref{sec:transfer} \\
$G_i(\nu)$ & transfer function of cell $i$ (unnormalised) & sr & \ref{sec:transfer} \\
$\langle t\rangle$ & first moment of $\psi$ (mean delay) & $\rgc$ & \ref{sec:transfer} \\
$\nu$, $\nu_k$, $\bar\nu_j$ & Fourier frequency; frequencies of bin $j$; bin mean & Hz, $t_g^{-1}$ & \ref{sec:likelihood} \\
$p_k$, $Z_j$ & power weight; binned model cross-spectrum (eq.~\ref{eq:estimator}) & -- & \ref{sec:likelihood} \\
$\tau(\nu)$, $\tau_j$ & soft minus hard lag (negative: soft lagging); binned lag & s, $t_g$ & \ref{sec:dilution} \\
$\sigma_{\tau,j}$, $\sigma_j$ & lag error; total error including dilution (eq.~\ref{eq:hlag}) & s, $t_g$ & \ref{sec:likelihood} \\
$R_s$, $R_h$, $R_E$, $R_{\rm ref}$ & reflected-to-continuum count-rate ratios in soft, hard, energy channel and reference bands & -- & \ref{sec:spectra} \\
$\Dil$, $\sigma_\Dil$ & dilution factor $R_s/(1+R_s)-R_h/(1+R_h)$ and its error & -- & \ref{sec:dilution} \\
$A$, $\gamma_c$, $\nu_0$ & amplitude, index and pivot of the continuum lag (eq.~\ref{eq:cont}) & s, --, Hz & \ref{sec:dilution} \\
$R_f$ & \textsc{relxill} reflection fraction (disc over infinity) & -- & \ref{sec:illum} \\
$I(\bm w)$, $\Iobs$, $\sigma_I$ & model and measured illumination fraction $R_f/(1+R_f)$; error & -- & \ref{sec:illum} \\
$\Imax$ & largest $I$ of any single cell ($0.564$) & -- & \ref{sec:illum} \\
$a$ (spin), $q$, $\log n_{\rm e}$ & dimensionless spin; emissivity index; disc density (spectral fits) & --, --, cm$^{-3}$ & \ref{sec:spectra} \\
\multicolumn{4}{@{}l}{\textit{Inference}}\\
$\Ham$, $\Hmag$, $\Hlag$, $\Hrad$ & Hamiltonian; smoothness, lag and illumination terms (eqs~\ref{eq:hlag}--\ref{eq:ham}) & -- & \ref{sec:hamiltonian} \\
$J$, $\kappa$ & smoothness coupling; normalisation $N^2/N_{\rm pairs}$ & -- & \ref{sec:hamiltonian} \\
$\alpha$, $\beta$ & weights of the lag and illumination terms ($=1$) & -- & \ref{sec:likelihood} \\
$F$, $\lambda_n$, $N_{\rm eff}$ & projected Fisher matrix; its eigenvalues; effective number of constrained parameters & -- & \ref{sec:info} \\
$\rho_{3D}$, $\rho_{Rz}$ & Pearson correlation of true and recovered maps (3D; azimuthally summed) & -- & \ref{sec:recovery} \\
$\Delta_{\rm cen}$ & distance between true and recovered centroids & $\rg$ & \ref{sec:recovery} \\
$\sigma$ & reflection $\phi\to-\phi$ about the observer's meridian plane & -- & \ref{sec:parity} \\
$h$ & lamppost height & $\rg$ & \ref{sec:lamppost} \\
\bottomrule
\end{tabular}
\end{table*}

\section{Validation on synthetic data}
\label{sec:validation}

\subsection{Recovery}
\label{sec:recovery}

Three axisymmetric input geometries are used: a compact Gaussian source on the axis at
$z=5\,\rg$ (width $2\,\rg$; lamppost-like), a column ($R<5$, $1<z<5\,\rg$) and a torus centred
at $R=8$, $z=3\,\rg$. For each we compute the lag-frequency spectrum with the exact transfer at
$\theta_{\rm obs}=30^\circ$ at four frequencies between $0.013$ and $0.045\,t_g^{-1}$ (the
reverberation band of the sources), with $R_s=0.6$, $R_h=0.15$, add 30 per cent Gaussian errors
and $I_{\rm obs}$ with $\sigma_I=0.05$, and fit 10 noise realisations (Table~\ref{tab:synthetic}, Fig.~\ref{fig:synthetic}).

The lags are reproduced to 5--9 per cent and the centroid to $2$--$4\,\rg$ for the
lamppost and column and $6\,\rg$ for the ring. The correlation between true and recovered
maps is $\rho_{3D}=0.51$, $0.39$ and $0.21$. These numbers reflect the information content
rather than a failure of the optimiser: $N_{\rm eff}=4.0$ in all cases, so the data fix
four combinations of 799 free weights and the rest of the map is the smoothest field
compatible with them. Averaging 20 solutions raises $\rho$ to $0.36$--$0.46$, as
expected when noise rather than bias dominates.

Fitting the same Schwarzschild data with flat-space transfer displaces the centroid by
$5$--$9\,\rg$, two to three times the error of the correct fit, although relativity changes
the mean delay of the input geometries by only 12--14 per cent: the geometry is sensitive to
the full delay distribution and to the illumination pattern, which light bending changes
much more than it changes $\langle t\rangle$. Exact transfer is therefore necessary for
the inferred geometry, not only for the lag amplitude. This reverses the conclusion of the
first version, where the morphology appeared insensitive to the Shapiro correction.

\begin{table*}
\centering
\caption{Synthetic recovery: medians and 16--84 percentiles over 10 noise realisations of 4 lag-frequency points (30 per cent errors) and $I_{\rm obs}$ ($\sigma_I=0.05$). $\Delta_{\rm cen}$: centroid offset; $\Delta^{\rm flat}_{\rm cen}$: the same data fitted with flat-space transfer. $\delta_{\rm GR}$: fractional increase of the mean delay $\langle t\rangle$ of the input geometry from flat space to Schwarzschild. $N_{\rm eff}$: data-constrained modes; $\rho(\bar{\bm w})$: correlation of the ensemble-mean map with the truth.}
\label{tab:synthetic}
\small
\begin{tabular}{lcccccccc}
\toprule
Geometry & $\rho_{3D}$ & $\rho_{Rz}$ & $\Delta_{\rm cen}$ ($r_g$) & $|\Delta\tau|/|\tau|$ & $\Delta^{\rm flat}_{\rm cen}$ ($r_g$) & $\delta_{\rm GR}$ & $N_{\rm eff}$ & $\rho(\bar{\bm w})$ \\
\midrule
lamppost & $0.51^{+0.03}_{-0.15}$ & $0.54^{+0.09}_{-0.11}$ & $3.5^{+0.8}_{-0.8}$ & $0.09^{+0.04}_{-0.06}$ & $7.7^{+0.9}_{-1.8}$ & 12\% & 4.0 & 0.46 \\
column & $0.39^{+0.08}_{-0.18}$ & $0.40^{+0.08}_{-0.17}$ & $2.0^{+0.6}_{-0.7}$ & $0.05^{+0.10}_{-0.03}$ & $5.3^{+1.6}_{-1.6}$ & 14\% & 4.0 & 0.36 \\
ring & $0.21^{+0.17}_{-0.19}$ & $0.22^{+0.16}_{-0.20}$ & $5.7^{+0.9}_{-3.9}$ & $0.06^{+0.17}_{-0.02}$ & $8.8^{+1.0}_{-1.7}$ & 12\% & 4.0 & 0.38 \\
\bottomrule
\end{tabular}
\end{table*}

\begin{figure*}
\centering
\begin{tikzpicture}
\begin{groupplot}[group style={group size=3 by 3, horizontal sep=6mm, vertical sep=12mm, xlabels at=edge bottom, ylabels at=edge left, yticklabels at=edge left},
  width=0.24\textwidth, height=0.24\textwidth, xmode=log, ymode=log, enlargelimits=false, axis on top, xtick={3,10,30}, ytick={0.1,1,10}, log ticks with fixed point, colormap name=hamcor, point meta min=0, point meta max=1,
  xlabel={$R\;[r_g]$}, ylabel={$z\;[r_g]$}, title style={font=\scriptsize, yshift=-1mm},
  tick label style={font=\scriptsize}, label style={font=\small}]
\nextgroupplot[title={lamppost: true}]
\addplot[matrix plot*, mesh/cols=10, point meta=explicit] table[x=x, y=y, meta=c] {figures/data/synthetic/true_lamppost.dat};
\nextgroupplot[title={lamppost: Schwarzschild fit}]
\addplot[matrix plot*, mesh/cols=10, point meta=explicit] table[x=x, y=y, meta=c] {figures/data/synthetic/rec_lamppost_gr.dat};
\nextgroupplot[title={lamppost: flat-space fit}]
\addplot[matrix plot*, mesh/cols=10, point meta=explicit] table[x=x, y=y, meta=c] {figures/data/synthetic/rec_lamppost_flat.dat};
\nextgroupplot[title={column: true}]
\addplot[matrix plot*, mesh/cols=10, point meta=explicit] table[x=x, y=y, meta=c] {figures/data/synthetic/true_column.dat};
\nextgroupplot[title={column: Schwarzschild fit}]
\addplot[matrix plot*, mesh/cols=10, point meta=explicit] table[x=x, y=y, meta=c] {figures/data/synthetic/rec_column_gr.dat};
\nextgroupplot[title={column: flat-space fit}]
\addplot[matrix plot*, mesh/cols=10, point meta=explicit] table[x=x, y=y, meta=c] {figures/data/synthetic/rec_column_flat.dat};
\nextgroupplot[title={ring: true}]
\addplot[matrix plot*, mesh/cols=10, point meta=explicit] table[x=x, y=y, meta=c] {figures/data/synthetic/true_ring.dat};
\nextgroupplot[title={ring: Schwarzschild fit}]
\addplot[matrix plot*, mesh/cols=10, point meta=explicit] table[x=x, y=y, meta=c] {figures/data/synthetic/rec_ring_gr.dat};
\nextgroupplot[title={ring: flat-space fit}, colorbar right, colorbar style={width=2mm, ytick={0,0.5,1}, ylabel={$\sum_\phi w_i/\max$}, ylabel style={font=\scriptsize}, tick label style={font=\scriptsize}}]
\addplot[matrix plot*, mesh/cols=10, point meta=explicit] table[x=x, y=y, meta=c] {figures/data/synthetic/rec_ring_flat.dat};
\end{groupplot}
\end{tikzpicture}
\caption{Synthetic recovery. Left: input geometries; centre: HAMCOR recovery of noisy lag-frequency data generated with the exact Schwarzschild transfer (one noise realisation); right: the same data fitted with flat-space light-travel times. All maps are normalised to their maximum.}
\label{fig:synthetic}
\end{figure*}

\subsection{Sensitivity to the prior}
\label{sec:robustness}

Fig.~\ref{fig:jscan} shows the recovery as a function of the smoothness coupling. Without
the prior ($J=0$) the solution is the sparsest fit of the data and the correlation drops
(from $0.54$ to $0.44$ for the lamppost and from $0.43$ to $0.31$ for the column); for
$0.1\le J\le10$ the correlation and the centroid change by less than $0.02$ and $0.5\,\rg$,
while $\chi^2$ of the data stays below the noise level. We adopt $J=1$.

Replacing the likelihood by the impulse-response construction of the first version with a
Gaussian, top-hat or uniform $\psi_{\rm obs}$ gives, for the lamppost, $\rho_{3D}=0.42$,
$0.43$ and $0.29$ and centroid offsets $1.8$, $1.8$ and $3.0\,\rg$; fitting the mean lag
alone gives $\rho_{3D}=0.35$. The lag-frequency likelihood is therefore at least as good as
the best of the earlier variants and does not depend on an arbitrary response shape.

\begin{figure}
\centering
\begin{tikzpicture}
\begin{groupplot}[group style={group size=1 by 2, vertical sep=4mm, xlabels at=edge bottom, xticklabels at=edge bottom}, width=0.95\columnwidth, height=0.42\columnwidth, xmode=log,
  xlabel={$J$}, tick label style={font=\scriptsize}, label style={font=\small},
  legend style={font=\scriptsize, draw=none, fill=none}, legend columns=3]
\nextgroupplot[ylabel={$\rho_{3D}$}, ymin=0, ymax=0.7, legend to name=jleg]
\addplot[thick, black, mark=*, mark size=1pt] table[x=J, y=rho] {figures/data/synthetic/Jscan_lamppost.dat}; \addlegendentry{lamppost}
\addplot[thick, blue!70!black, dashed, mark=*, mark size=1pt] table[x=J, y=rho] {figures/data/synthetic/Jscan_column.dat}; \addlegendentry{column}
\addplot[thick, red!70!black, dotted, mark=*, mark size=1pt] table[x=J, y=rho] {figures/data/synthetic/Jscan_ring.dat}; \addlegendentry{ring}
\nextgroupplot[ylabel={$\Delta_{\rm cen}$ [$r_g$]}, ymin=0]
\addplot[thick, black, mark=*, mark size=1pt] table[x=J, y=dcen] {figures/data/synthetic/Jscan_lamppost.dat};
\addplot[thick, blue!70!black, dashed, mark=*, mark size=1pt] table[x=J, y=dcen] {figures/data/synthetic/Jscan_column.dat};
\addplot[thick, red!70!black, dotted, mark=*, mark size=1pt] table[x=J, y=dcen] {figures/data/synthetic/Jscan_ring.dat};
\end{groupplot}
\node at (current bounding box.north) [above] {\pgfplotslegendfromname{jleg}};
\end{tikzpicture}
\caption{Sensitivity to the smoothness coupling $J$ (the point plotted at the left edge is $J=0$). Top: correlation between true and recovered maps; bottom: centroid offset.}
\label{fig:jscan}
\end{figure}

\subsection{What the data cannot see: azimuthal parity}
\label{sec:parity}

The reflection $\sigma:(R,z,\phi)\to(R,z,-\phi)$ about the plane containing the spin axis
and the line of sight is an isometry that fixes the hole. Schwarzschild spacetime is
spherically symmetric, so geodesics, solid angles, captured and escaping fractions and
delays are all invariant: $\Omega_{\sigma(i)\sigma(k)}=\Omega_{ik}$,
$L_{\sigma(i)\sigma(k)}=L_{ik}$, $E_{\sigma(i)}=E_i$. Re-indexing the sums in
equations~(\ref{eq:psitilde}) and (\ref{eq:illum}) gives $\tilde\psi(\nu;\bm w\circ\sigma)=
\tilde\psi(\nu;\bm w)$ and $I(\bm w\circ\sigma)=I(\bm w)$ for every $\bm w$; the
neighbour graph, and hence $\Hmag$, is also invariant. The part of the emissivity that is
odd under $\sigma$ is therefore unobservable with any combination of lags, transfer
function and reflection fraction, at any resolution. Numerically, the response of the lags
to the odd part of a map is $1.5\times10^{-4}$ of that to the even part (ray-sampling noise)
and that of $I$ is zero; an input with 50 per cent odd power is recovered with an odd
fraction of $10^{-8}$. The axisymmetric test geometries of Table~\ref{tab:synthetic} have no odd part; the ring is
the hardest of them because its emission spans a wide range of delays that a smooth extended
field reproduces equally well, not because of this degeneracy. The first version attributed its
poor recovery to parity. Frame dragging
($a\neq0$) and the polarization angle, both odd under $\sigma$, would break the degeneracy
(Section~\ref{sec:limitations}).

\section{Tests of the ray tracing}
\label{app:tests}

The ray-tracing module is covered by unit tests run with the pipeline: (i) with the force
term switched off, the fraction of rays from $h=5\,\rg$ reaching an annulus
$6\le r\le100$ equals the analytic solid angle to $10^{-4}$, the stored times equal the
Euclidean distances to $10^{-8}$, and $T(\bm x)=-\bm x\cdot\nhat$ to $10^{-6}$; (ii) in
Schwarzschild, the captured fraction at $r=5$ equals the analytic capture cone
($0.2034$); (iii) at $h=300\,\rg$ the disc fraction agrees with the flat value within the
sampling noise; (iv) for a source behind the hole ($\bm x=(-8,0,0.5)$), every reflected ray
arrives after the direct one (minimum lag $1.1\,\rgc$). On the production grid, fewer than
$2\times10^{-2}$ per cent of rays have $L<-1\,\rgc$ before the causality clip of
equation~(\ref{eq:lagmatrix}); these residuals come from evaluating $T$ at the centre of
the disc element rather than at the hit point.

\section{Weak-field Shapiro delay}
\label{app:weak}

To first order in $M/r$, the extra coordinate time along a straight segment between points
at radii $r_A$ and $r_B$ separated by $\ell$ is
\begin{equation}
  S(A,B)=2\ln\frac{r_A+r_B+\ell}{r_A+r_B-\ell}
  \label{eq:shapiro}
\end{equation}
\citep{Shapiro1964}, exact (to this order) both for segments with an interior point of
closest approach to the hole (V-shaped legs) and for monotonic ones. The first version of
this work used a closest-approach formula that applies only to V-shaped legs and returned
a fixed cap of $20\,\rgc$ for monotonic legs, which are the majority, and projected the
observer leg with the wrong sign; together these produced the quoted average correction
of $\sim79$ per cent. With equation~(\ref{eq:shapiro}) and $S_\infty(\bm x)=-2\ln[r(1+\cos\vartheta)]$
for the leg to infinity, the solid-angle-weighted Shapiro correction over the grid is
12--13 per cent, consistent with the 12--14 per cent change of the mean delay obtained with
the exact transfer (Table~\ref{tab:synthetic}). The weak-field model ignores light bending and
capture; it is retained as a systematic test (\textit{weak-field} in Table~\ref{tab:fits}).

\section{Lag-energy spectra, spectral residuals and map scatter}
\label{app:lagenergy}

The lag-energy spectrum in the reverberation band is predicted from the HAMCOR and
lamppost solutions using, for each channel, the reflected-to-continuum ratio $R_E$ from the
spectral fit and the complementary reference band \citep{Uttley2014}:
\begin{equation}
  \tau(E)=-\frac{\arg\sum_kp_k[1+R_{\rm ref}\tilde\psi(\nu_k)]^*[1+R_E\tilde\psi(\nu_k)]}{2\pi\bar\nu},
\end{equation}
with a free additive offset. Fig.~\ref{fig:lagenergy} compares the predictions with the
data; Fig.~\ref{fig:spectra} shows the spectral residuals and Fig.~\ref{fig:mapscv} the
bootstrap scatter of the emissivity maps.

\begin{figure*}
\centering
\begin{tikzpicture}
\begin{groupplot}[group style={group size=4 by 2, horizontal sep=11mm, vertical sep=13mm, xlabels at=edge bottom},
  width=0.27\textwidth, height=0.22\textwidth, xmode=log, xmin=0.3, xmax=10,
  xtick={0.5,1,2,5}, log ticks with fixed point, xlabel={$E$ [keV]},
  tick label style={font=\scriptsize}, label style={font=\small},
  title style={font=\scriptsize, yshift=-1mm}]
\nextgroupplot[title={Mrk 335 (2006)}, ylabel={lag [s]}]
\addplot[only marks, mark=*, mark size=1pt, black, error bars/.cd, x dir=both, x explicit, y dir=both, y explicit] table[x=E, y=lag, x error=dE, y error=sig] {figures/data/mrk335_2006/lagenergy.dat};
\addplot[thick, blue!70!black] table[x=E, y=hamcor] {figures/data/mrk335_2006/lagenergy_model.dat};
\addplot[thick, dashed, red!70!black] table[x=E, y=lamppost] {figures/data/mrk335_2006/lagenergy_model.dat};
\nextgroupplot[title={Mrk 335 (2009)}]
\addplot[only marks, mark=*, mark size=1pt, black, error bars/.cd, x dir=both, x explicit, y dir=both, y explicit] table[x=E, y=lag, x error=dE, y error=sig] {figures/data/mrk335_2009/lagenergy.dat};
\addplot[thick, blue!70!black] table[x=E, y=hamcor] {figures/data/mrk335_2009/lagenergy_model.dat};
\addplot[thick, dashed, red!70!black] table[x=E, y=lamppost] {figures/data/mrk335_2009/lagenergy_model.dat};
\nextgroupplot[title={Mrk 335 (2015)}]
\addplot[only marks, mark=*, mark size=1pt, black, error bars/.cd, x dir=both, x explicit, y dir=both, y explicit] table[x=E, y=lag, x error=dE, y error=sig] {figures/data/mrk335_2015/lagenergy.dat};
\addplot[thick, blue!70!black] table[x=E, y=hamcor] {figures/data/mrk335_2015/lagenergy_model.dat};
\addplot[thick, dashed, red!70!black] table[x=E, y=lamppost] {figures/data/mrk335_2015/lagenergy_model.dat};
\nextgroupplot[title={Mrk 335 (2018)}]
\addplot[only marks, mark=*, mark size=1pt, black, error bars/.cd, x dir=both, x explicit, y dir=both, y explicit] table[x=E, y=lag, x error=dE, y error=sig] {figures/data/mrk335_2018/lagenergy.dat};
\addplot[thick, blue!70!black] table[x=E, y=hamcor] {figures/data/mrk335_2018/lagenergy_model.dat};
\addplot[thick, dashed, red!70!black] table[x=E, y=lamppost] {figures/data/mrk335_2018/lagenergy_model.dat};
\nextgroupplot[title={Mrk 335 (2019)}, ylabel={lag [s]}]
\addplot[only marks, mark=*, mark size=1pt, black, error bars/.cd, x dir=both, x explicit, y dir=both, y explicit] table[x=E, y=lag, x error=dE, y error=sig] {figures/data/mrk335_2019/lagenergy.dat};
\addplot[thick, blue!70!black] table[x=E, y=hamcor] {figures/data/mrk335_2019/lagenergy_model.dat};
\addplot[thick, dashed, red!70!black] table[x=E, y=lamppost] {figures/data/mrk335_2019/lagenergy_model.dat};
\nextgroupplot[title={1H 0707$-$495}]
\addplot[only marks, mark=*, mark size=1pt, black, error bars/.cd, x dir=both, x explicit, y dir=both, y explicit] table[x=E, y=lag, x error=dE, y error=sig] {figures/data/1h0707/lagenergy.dat};
\addplot[thick, blue!70!black] table[x=E, y=hamcor] {figures/data/1h0707/lagenergy_model.dat};
\addplot[thick, dashed, red!70!black] table[x=E, y=lamppost] {figures/data/1h0707/lagenergy_model.dat};
\nextgroupplot[title={IRAS 13224$-$3809}]
\addplot[only marks, mark=*, mark size=1pt, black, error bars/.cd, x dir=both, x explicit, y dir=both, y explicit] table[x=E, y=lag, x error=dE, y error=sig] {figures/data/iras13224/lagenergy.dat};
\addplot[thick, blue!70!black] table[x=E, y=hamcor] {figures/data/iras13224/lagenergy_model.dat};
\addplot[thick, dashed, red!70!black] table[x=E, y=lamppost] {figures/data/iras13224/lagenergy_model.dat};
\nextgroupplot[title={MCG$-$6$-$30$-$15}]
\addplot[only marks, mark=*, mark size=1pt, black, error bars/.cd, x dir=both, x explicit, y dir=both, y explicit] table[x=E, y=lag, x error=dE, y error=sig] {figures/data/mcg6/lagenergy.dat};
\addplot[thick, blue!70!black] table[x=E, y=hamcor] {figures/data/mcg6/lagenergy_model.dat};
\addplot[thick, dashed, red!70!black] table[x=E, y=lamppost] {figures/data/mcg6/lagenergy_model.dat};
\end{groupplot}
\end{tikzpicture}
\caption{Lag-energy spectra in the reverberation band (not used in any fit). Blue: prediction of the HAMCOR solution; red dashed: best lamppost; both up to a common additive offset (the only free parameter).}
\label{fig:lagenergy}
\end{figure*}
\begin{figure*}
\centering
\begin{tikzpicture}
\begin{groupplot}[group style={group size=4 by 2, horizontal sep=10mm, vertical sep=13mm, xlabels at=edge bottom},
  width=0.27\textwidth, height=0.20\textwidth, xmode=log, xmin=0.3, xmax=10, ymin=0.8, ymax=1.2,
  xtick={0.5,1,2,5}, log ticks with fixed point, xlabel={$E$ [keV]},
  tick label style={font=\scriptsize}, label style={font=\small},
  title style={font=\scriptsize, yshift=-1mm}, extra y ticks={1}, extra y tick labels={},
  extra y tick style={grid=major, major grid style={densely dotted, gray}}]
\nextgroupplot[title={Mrk 335 (2006)}, ylabel={data/model}]
\addplot[only marks, mark=*, mark size=0.5pt, black, error bars/.cd, y dir=both, y explicit, error bar style={gray}] table[x=E, y=r, y error=dr] {figures/data/mrk335_2006/spec_ratio.dat};
\nextgroupplot[title={Mrk 335 (2009)}]
\addplot[only marks, mark=*, mark size=0.5pt, black, error bars/.cd, y dir=both, y explicit, error bar style={gray}] table[x=E, y=r, y error=dr] {figures/data/mrk335_2009/spec_ratio.dat};
\nextgroupplot[title={Mrk 335 (2015)}]
\addplot[only marks, mark=*, mark size=0.5pt, black, error bars/.cd, y dir=both, y explicit, error bar style={gray}] table[x=E, y=r, y error=dr] {figures/data/mrk335_2015/spec_ratio.dat};
\nextgroupplot[title={Mrk 335 (2018)}]
\addplot[only marks, mark=*, mark size=0.5pt, black, error bars/.cd, y dir=both, y explicit, error bar style={gray}] table[x=E, y=r, y error=dr] {figures/data/mrk335_2018/spec_ratio.dat};
\nextgroupplot[title={Mrk 335 (2019)}, ylabel={data/model}]
\addplot[only marks, mark=*, mark size=0.5pt, black, error bars/.cd, y dir=both, y explicit, error bar style={gray}] table[x=E, y=r, y error=dr] {figures/data/mrk335_2019/spec_ratio.dat};
\nextgroupplot[title={1H 0707$-$495}]
\addplot[only marks, mark=*, mark size=0.5pt, black, error bars/.cd, y dir=both, y explicit, error bar style={gray}] table[x=E, y=r, y error=dr] {figures/data/1h0707/spec_ratio.dat};
\nextgroupplot[title={IRAS 13224$-$3809}]
\addplot[only marks, mark=*, mark size=0.5pt, black, error bars/.cd, y dir=both, y explicit, error bar style={gray}] table[x=E, y=r, y error=dr] {figures/data/iras13224/spec_ratio.dat};
\nextgroupplot[title={MCG$-$6$-$30$-$15}]
\addplot[only marks, mark=*, mark size=0.5pt, black, error bars/.cd, y dir=both, y explicit, error bar style={gray}] table[x=E, y=r, y error=dr] {figures/data/mcg6/spec_ratio.dat};
\end{groupplot}
\end{tikzpicture}
\caption{EPIC-pn spectral residuals (data/model) for the reflection fits of Table~\ref{tab:spectral} ($a=0$, $R_{\rm in}=6\,r_g$).}
\label{fig:spectra}
\end{figure*}
\begin{figure*}
\centering
\begin{tikzpicture}
\begin{groupplot}[group style={group size=4 by 2, horizontal sep=6mm, vertical sep=12mm, xlabels at=edge bottom, ylabels at=edge left, yticklabels at=edge left},
  width=0.235\textwidth, height=0.235\textwidth, xmode=log, ymode=log, enlargelimits=false, axis on top, xtick={3,10,30}, ytick={0.1,1,10}, log ticks with fixed point, colormap name=hamcor, point meta min=0, point meta max=2,
  xlabel={$R\;[r_g]$}, ylabel={$z\;[r_g]$}, title style={font=\scriptsize, yshift=-1mm},
  tick label style={font=\scriptsize}, label style={font=\small}]
\nextgroupplot[title={Mrk 335 (2006)}]
\addplot[matrix plot*, mesh/cols=10, point meta=explicit] table[x=x, y=y, meta=c] {figures/data/mrk335_2006/map_bootcv.dat};
\nextgroupplot[title={Mrk 335 (2009)}]
\addplot[matrix plot*, mesh/cols=10, point meta=explicit] table[x=x, y=y, meta=c] {figures/data/mrk335_2009/map_bootcv.dat};
\nextgroupplot[title={Mrk 335 (2015)}]
\addplot[matrix plot*, mesh/cols=10, point meta=explicit] table[x=x, y=y, meta=c] {figures/data/mrk335_2015/map_bootcv.dat};
\nextgroupplot[title={Mrk 335 (2018)}]
\addplot[matrix plot*, mesh/cols=10, point meta=explicit] table[x=x, y=y, meta=c] {figures/data/mrk335_2018/map_bootcv.dat};
\nextgroupplot[title={Mrk 335 (2019)}]
\addplot[matrix plot*, mesh/cols=10, point meta=explicit] table[x=x, y=y, meta=c] {figures/data/mrk335_2019/map_bootcv.dat};
\nextgroupplot[title={1H 0707$-$495}]
\addplot[matrix plot*, mesh/cols=10, point meta=explicit] table[x=x, y=y, meta=c] {figures/data/1h0707/map_bootcv.dat};
\nextgroupplot[title={IRAS 13224$-$3809}]
\addplot[matrix plot*, mesh/cols=10, point meta=explicit] table[x=x, y=y, meta=c] {figures/data/iras13224/map_bootcv.dat};
\nextgroupplot[title={MCG$-$6$-$30$-$15}, colorbar right, colorbar style={width=2mm, ytick={0,1,2}, ylabel={$\sigma/\bar w$}, ylabel style={font=\scriptsize}, tick label style={font=\scriptsize}}]
\addplot[matrix plot*, mesh/cols=10, point meta=explicit] table[x=x, y=y, meta=c] {figures/data/mcg6/map_bootcv.dat};
\end{groupplot}
\end{tikzpicture}
\caption{Bootstrap coefficient of variation (standard deviation over mean) of the maps of Fig.~\ref{fig:maps}.}
\label{fig:mapscv}
\end{figure*}

\section{Additional tables and figures}
\label{app:extra}

\begin{table*}
\centering
\caption{\textit{XMM-Newton} EPIC-pn observations. $T$: length of the light curves after flare filtering; $\langle r_s\rangle$, $\langle r_h\rangle$: background-subtracted, live-time corrected count rates (ct\,s$^{-1}$) in 0.3--1 and 1.5--4\,keV.}
\label{tab:obs}
\small
\begin{tabular}{lllcccc}
\toprule
Source & ObsID & Date & pn mode & $T$ (ks) & $\langle r_s\rangle$ & $\langle r_h\rangle$ \\
\midrule
Mrk 335 (2006) & 0306870101 & 2006-01-03 & Small W. & 132.8 & 18.20 & 3.46 \\
Mrk 335 (2009) & 0600540601 & 2009-06-11 & Full W. & 130.4 & 1.99 & 0.61 \\
Mrk 335 (2015) & 0741280201 & 2015-12-30 & Full W. & 137.4 & 0.69 & 0.19 \\
Mrk 335 (2018) & 0780500301 & 2018-07-11 & Full W. & 111.3 & 0.60 & 0.08 \\
Mrk 335 (2019) & 0831790601 & 2019-01-08 & Full W. & 114.6 & 0.47 & 0.07 \\
1H 0707$-$495 & 0511580101 & 2008-01-29 & Large W. & 121.6 & 3.15 & 0.27 \\
IRAS 13224$-$3809 & 0780560101 & 2016-07-09 & Large W. & 52.1 & 2.29 & 0.12 \\
MCG$-$6$-$30$-$15 & 0029740701 & 2001-08-02 & Small W. & 127.4 & 15.77 & 6.97 \\
\bottomrule
\end{tabular}
\end{table*}

\begin{table*}
\centering
\caption{Mrk~335: comparison with previous work.}
\label{tab:literature}
\small
\begin{tabular}{@{}p{0.15\textwidth}p{0.17\textwidth}p{0.2\textwidth}p{0.42\textwidth}@{}}
\toprule
Reference & Data & Method & Result \\
\midrule
\citet{Kara2013} & \xmm{} 2006 & lag-frequency, lag-energy & soft lag $-155\pm57$\,s at $\sim2\times10^{-4}$\,Hz; Fe\,K lag $\sim150$\,s \\
\citet{Parker2014} & NuSTAR+\textit{Swift} 2013 & lamppost spectra & $h<2.3$ to $5.2^{+2.2}_{-0.8}\,r_g$; $R_f=2.7$ to $>8$; $a>0.9$ \\
\citet{WilkinsGallo2015} & \xmm{}, \textit{Suzaku} 2006--2013 & emissivity profiles & extent $26^{+10}_{-7}\,r_g$ (2006), $\lesssim12\,r_g$ (2009), $\lesssim5\,r_g$ (low state) \\
\citet{Chainakun2016} & \xmm{} 2006 & lamppost spectra + lag-energy & $h=3.0^{+0.1}_{-0.3}\,r_g$ \\
This work & \xmm{} 2006 & HAMCOR, lags & $R_c=9.1$, $z_c=3.1\,r_g$ (Table~\ref{tab:fits}); lamppost $h=2.6\,r_g$ \\
This work & \xmm{} 2009 & HAMCOR, lags $+I_{\rm obs}$ & $R_c=11.7$, $z_c=6.9\,r_g$ (Table~\ref{tab:fits}); lamppost $h=31.0\,r_g$ \\
This work & \xmm{} 2015 & HAMCOR, lags & $R_c=8.1$, $z_c=4.1\,r_g$ (Table~\ref{tab:fits}); lamppost $h=12.8\,r_g$ \\
This work & \xmm{} 2018 & HAMCOR, lags & $R_c=8.6$, $z_c=5.0\,r_g$ (Table~\ref{tab:fits}); lamppost $h=4.6\,r_g$ \\
This work & \xmm{} 2019 & HAMCOR, lags & $R_c=8.2$, $z_c=2.9\,r_g$ (Table~\ref{tab:fits}); lamppost $h=7.1\,r_g$ \\
\bottomrule
\end{tabular}
\end{table*}

\begin{figure*}
\centering
\begin{tikzpicture}
\begin{groupplot}[group style={group size=4 by 1, horizontal sep=6mm, vertical sep=12mm, xlabels at=edge bottom, ylabels at=edge left, yticklabels at=edge left},
  width=0.235\textwidth, height=0.235\textwidth, xmode=log, ymode=log, enlargelimits=false, axis on top, xtick={3,10,30}, ytick={0.1,1,10}, log ticks with fixed point, colormap name=hamdiv, point meta min=-1, point meta max=1,
  xlabel={$R\;[r_g]$}, ylabel={$z\;[r_g]$}, title style={font=\scriptsize, yshift=-1mm},
  tick label style={font=\scriptsize}, label style={font=\small}]
\nextgroupplot[title={mode 1: $\lambda=2.8\times10^{4}$}]
\addplot[matrix plot*, mesh/cols=10, point meta=explicit] table[x=x, y=y, meta=c] {figures/data/1h0707/mode1.dat};
\nextgroupplot[title={mode 2: $\lambda=1.1\times10^{3}$}]
\addplot[matrix plot*, mesh/cols=10, point meta=explicit] table[x=x, y=y, meta=c] {figures/data/1h0707/mode2.dat};
\nextgroupplot[title={mode 3: $\lambda=4.3\times10^{1}$}]
\addplot[matrix plot*, mesh/cols=10, point meta=explicit] table[x=x, y=y, meta=c] {figures/data/1h0707/mode3.dat};
\nextgroupplot[title={mode 4: $\lambda=2.7\times10^{0}$}, colorbar right, colorbar style={width=2mm, ytick={-1,0,1}, ylabel={mode (a.u.)}, ylabel style={font=\scriptsize}, tick label style={font=\scriptsize}}]
\addplot[matrix plot*, mesh/cols=10, point meta=explicit] table[x=x, y=y, meta=c] {figures/data/1h0707/mode4.dat};
\end{groupplot}
\end{tikzpicture}
\caption{The four data-constrained directions of the emissivity for 1H 0707$-$495 (right singular vectors of the noise-weighted Jacobian of the observables, summed over $\phi$ and normalised; $\lambda$: Fisher eigenvalue; $N_{\rm eff}=4.1$). Correlations of each mode with the per-cell mean delay, radius, height and illumination: mode 1: $\langle t\rangle_i$ $-0.86$, $r_i$ $+0.07$, $z_i$ $-0.73$, $I_i$ $+0.36$; mode 2: $\langle t\rangle_i$ $+0.40$, $r_i$ $+0.23$, $z_i$ $+0.55$, $I_i$ $+0.10$; mode 3: $\langle t\rangle_i$ $+0.11$, $r_i$ $-0.01$, $z_i$ $+0.08$, $I_i$ $+0.03$; mode 4: $\langle t\rangle_i$ $+0.07$, $r_i$ $+0.41$, $z_i$ $+0.18$, $I_i$ $+0.16$.}
\label{fig:modes}
\end{figure*}
\begin{table*}
\centering
\caption{Soft (0.3--1\,keV) minus hard (1.5--4\,keV) lags averaged over the reverberation band (negative: soft lagging). $\tau/\mathcal{D}$ is the dilution-corrected estimate of $-\langle t\rangle$ in the low-frequency limit (equation~\ref{eq:lowf}); for the 2019 epoch of Mrk~335 the band lag is dominated by the continuum (hard) lag.}
\label{tab:lags}
\small
\begin{tabular}{lcccccc}
\toprule
Source & $M_{\rm BH}$ (M$_\odot$) & band ($10^{-4}$\,Hz) & $N_{\rm seg}$ & $\tau$ (s) & $\tau$ ($r_g/c$) & $\tau/\mathcal{D}$ ($r_g/c$) \\
\midrule
Mrk 335 (2006) & 1.3$\times10^{7}$ & 2--7 & 5 & $-73\pm32$ & $-1.1\pm0.5$ & $-7.1$ \\
Mrk 335 (2009) & 1.3$\times10^{7}$ & 2--7 & 9 & $-87\pm109$ & $-1.4\pm1.7$ & $-3.5$ \\
Mrk 335 (2015) & 1.3$\times10^{7}$ & 2--7 & 5 & $-563\pm372$ & $-8.8\pm5.8$ & $-22.6$ \\
Mrk 335 (2018) & 1.3$\times10^{7}$ & 2--7 & 4 & $-635\pm284$ & $-9.9\pm4.4$ & $-31.5$ \\
Mrk 335 (2019) & 1.3$\times10^{7}$ & 2--7 & 4 & $+734\pm171$ & $+11.5\pm2.7$ & $+41.7$ \\
1H 0707$-$495 & 2$\times10^{6}$ & 6--30 & 19 & $-19\pm6$ & $-1.9\pm0.6$ & $-4.8$ \\
IRAS 13224$-$3809 & 1.9$\times10^{6}$ & 6--30 & 2 & $-13\pm12$ & $-1.4\pm1.3$ & $-3.8$ \\
MCG$-$6$-$30$-$15 & 1.6$\times10^{6}$ & 6--20 & 19 & $-15\pm11$ & $-1.9\pm1.4$ & $-8.1$ \\
\bottomrule
\end{tabular}
\end{table*}

\begin{figure*}
\centering
\begin{tikzpicture}
\begin{groupplot}[group style={group size=4 by 2, horizontal sep=11mm, vertical sep=13mm, xlabels at=edge bottom},
  width=0.27\textwidth, height=0.24\textwidth, xmode=log, xmin=3e-5, xmax=6e-3,
  xlabel={$f$ [Hz]}, tick label style={font=\scriptsize}, label style={font=\small},
  title style={font=\scriptsize, yshift=-1mm}, extra y ticks={0}, extra y tick labels={},
  extra y tick style={grid=major, major grid style={densely dotted, gray}}]
\nextgroupplot[title={Mrk 335 (2006)}, ylabel={$\tau$ [s]}, restrict y to domain*=-3000:3000]
\addplot[only marks, mark=o, mark size=0.9pt, gray, error bars/.cd, y dir=both, y explicit, error bar style={gray!60}] table[x=f, y=tau, y error=sig] {figures/data/mrk335_2006/lagfull.dat};
\addplot[only marks, mark=*, mark size=1.1pt, black, error bars/.cd, y dir=both, y explicit] table[x=f, y=tau, y error=sig] {figures/data/mrk335_2006/lagfit.dat};
\addplot[thick, blue!70!black] table[x=f, y=tau] {figures/data/mrk335_2006/lagmodel.dat};
\addplot[thin, dashed, red!70!black] table[x=f, y=rev] {figures/data/mrk335_2006/lagmodel.dat};
\nextgroupplot[title={Mrk 335 (2009)}, restrict y to domain*=-3000:3000]
\addplot[only marks, mark=o, mark size=0.9pt, gray, error bars/.cd, y dir=both, y explicit, error bar style={gray!60}] table[x=f, y=tau, y error=sig] {figures/data/mrk335_2009/lagfull.dat};
\addplot[only marks, mark=*, mark size=1.1pt, black, error bars/.cd, y dir=both, y explicit] table[x=f, y=tau, y error=sig] {figures/data/mrk335_2009/lagfit.dat};
\addplot[thick, blue!70!black] table[x=f, y=tau] {figures/data/mrk335_2009/lagmodel.dat};
\addplot[thin, dashed, red!70!black] table[x=f, y=rev] {figures/data/mrk335_2009/lagmodel.dat};
\nextgroupplot[title={Mrk 335 (2015)}, restrict y to domain*=-3000:3000]
\addplot[only marks, mark=o, mark size=0.9pt, gray, error bars/.cd, y dir=both, y explicit, error bar style={gray!60}] table[x=f, y=tau, y error=sig] {figures/data/mrk335_2015/lagfull.dat};
\addplot[only marks, mark=*, mark size=1.1pt, black, error bars/.cd, y dir=both, y explicit] table[x=f, y=tau, y error=sig] {figures/data/mrk335_2015/lagfit.dat};
\addplot[thick, blue!70!black] table[x=f, y=tau] {figures/data/mrk335_2015/lagmodel.dat};
\addplot[thin, dashed, red!70!black] table[x=f, y=rev] {figures/data/mrk335_2015/lagmodel.dat};
\nextgroupplot[title={Mrk 335 (2018)}, restrict y to domain*=-3000:3000]
\addplot[only marks, mark=o, mark size=0.9pt, gray, error bars/.cd, y dir=both, y explicit, error bar style={gray!60}] table[x=f, y=tau, y error=sig] {figures/data/mrk335_2018/lagfull.dat};
\addplot[only marks, mark=*, mark size=1.1pt, black, error bars/.cd, y dir=both, y explicit] table[x=f, y=tau, y error=sig] {figures/data/mrk335_2018/lagfit.dat};
\addplot[thick, blue!70!black] table[x=f, y=tau] {figures/data/mrk335_2018/lagmodel.dat};
\addplot[thin, dashed, red!70!black] table[x=f, y=rev] {figures/data/mrk335_2018/lagmodel.dat};
\nextgroupplot[title={Mrk 335 (2019)}, ylabel={$\tau$ [s]}, restrict y to domain*=-3000:3000]
\addplot[only marks, mark=o, mark size=0.9pt, gray, error bars/.cd, y dir=both, y explicit, error bar style={gray!60}] table[x=f, y=tau, y error=sig] {figures/data/mrk335_2019/lagfull.dat};
\addplot[only marks, mark=*, mark size=1.1pt, black, error bars/.cd, y dir=both, y explicit] table[x=f, y=tau, y error=sig] {figures/data/mrk335_2019/lagfit.dat};
\addplot[thick, blue!70!black] table[x=f, y=tau] {figures/data/mrk335_2019/lagmodel.dat};
\addplot[thin, dashed, red!70!black] table[x=f, y=rev] {figures/data/mrk335_2019/lagmodel.dat};
\nextgroupplot[title={1H 0707$-$495}, restrict y to domain*=-3000:3000]
\addplot[only marks, mark=o, mark size=0.9pt, gray, error bars/.cd, y dir=both, y explicit, error bar style={gray!60}] table[x=f, y=tau, y error=sig] {figures/data/1h0707/lagfull.dat};
\addplot[only marks, mark=*, mark size=1.1pt, black, error bars/.cd, y dir=both, y explicit] table[x=f, y=tau, y error=sig] {figures/data/1h0707/lagfit.dat};
\addplot[thick, blue!70!black] table[x=f, y=tau] {figures/data/1h0707/lagmodel.dat};
\addplot[thin, dashed, red!70!black] table[x=f, y=rev] {figures/data/1h0707/lagmodel.dat};
\nextgroupplot[title={IRAS 13224$-$3809}, restrict y to domain*=-3000:3000]
\addplot[only marks, mark=o, mark size=0.9pt, gray, error bars/.cd, y dir=both, y explicit, error bar style={gray!60}] table[x=f, y=tau, y error=sig] {figures/data/iras13224/lagfull.dat};
\addplot[only marks, mark=*, mark size=1.1pt, black, error bars/.cd, y dir=both, y explicit] table[x=f, y=tau, y error=sig] {figures/data/iras13224/lagfit.dat};
\addplot[thick, blue!70!black] table[x=f, y=tau] {figures/data/iras13224/lagmodel.dat};
\addplot[thin, dashed, red!70!black] table[x=f, y=rev] {figures/data/iras13224/lagmodel.dat};
\nextgroupplot[title={MCG$-$6$-$30$-$15}, restrict y to domain*=-3000:3000]
\addplot[only marks, mark=o, mark size=0.9pt, gray, error bars/.cd, y dir=both, y explicit, error bar style={gray!60}] table[x=f, y=tau, y error=sig] {figures/data/mcg6/lagfull.dat};
\addplot[only marks, mark=*, mark size=1.1pt, black, error bars/.cd, y dir=both, y explicit] table[x=f, y=tau, y error=sig] {figures/data/mcg6/lagfit.dat};
\addplot[thick, blue!70!black] table[x=f, y=tau] {figures/data/mcg6/lagmodel.dat};
\addplot[thin, dashed, red!70!black] table[x=f, y=rev] {figures/data/mcg6/lagmodel.dat};
\end{groupplot}
\end{tikzpicture}
\caption{Soft (0.3--1\,keV) minus hard (1.5--4\,keV) lag-frequency spectra. Grey: all frequencies; black: bins in the reverberation band used in the fit; blue: HAMCOR model evaluated through the same cross-spectral estimator (reverberation plus profiled continuum lag); red dashed: reverberation component alone. Negative lags: soft band lagging.}
\label{fig:lagfreq}
\end{figure*}
\begin{figure}
\centering
\begin{tikzpicture}
\begin{axis}[width=0.95\columnwidth, height=0.55\columnwidth, xmode=log, xmin=2.5, xmax=40,
  xtick={3,5,10,20,40}, log ticks with fixed point, xlabel={$h$ [$r_g$]},
  ylabel={$I=R_f/(1+R_f)$}, ymin=0, ymax=0.6, tick label style={font=\scriptsize},
  label style={font=\small}, legend style={font=\scriptsize, draw=none, fill=none}, legend pos=south east]
\addplot[thick, black] table[x=h, y=I_gr] {figures/data/lamppost_I.dat}; \addlegendentry{Schwarzschild, ray traced}
\addplot[thick, dashed, gray] table[x=h, y=I_flat] {figures/data/lamppost_I.dat}; \addlegendentry{flat space}
\addplot[densely dotted, red!70!black, thick, domain=2.5:40] {0.564}; \addlegendentry{$I_{\max}$ (any cell)}
\end{axis}
\end{tikzpicture}
\par\smallskip
\begin{tikzpicture}
\begin{axis}[width=0.62\columnwidth, height=0.62\columnwidth, xmode=log, ymode=log, enlargelimits=false, axis on top, xtick={3,10,30}, ytick={0.1,1,10}, log ticks with fixed point, colormap name=hamcor, point meta min=0.1, point meta max=0.57,
  xlabel={$R\;[r_g]$}, ylabel={$z\;[r_g]$}, colorbar, colorbar style={width=2mm, ylabel={$I$}, tick label style={font=\scriptsize}}, tick label style={font=\scriptsize}, label style={font=\small}]
\addplot[matrix plot*, mesh/cols=10, point meta=explicit] table[x=x, y=y, meta=I] {figures/data/Imap.dat};
\end{axis}
\end{tikzpicture}
\caption{Illumination fraction (photons reaching the disc over photons reaching the disc or infinity) for a disc between $6$ and $100\,r_g$. Top: on-axis point source; bottom: single coronal cells of the HAMCOR grid (exact Schwarzschild transfer). No isotropic source can exceed $I_{\max}=0.56$ ($R_f\simeq1.3$).}
\label{fig:illum}
\end{figure}
\begin{figure*}
\centering
\begin{tikzpicture}
\begin{groupplot}[group style={group size=4 by 2, horizontal sep=6mm, vertical sep=12mm, xlabels at=edge bottom, ylabels at=edge left, yticklabels at=edge left},
  width=0.235\textwidth, height=0.235\textwidth, xmode=log, ymode=log, enlargelimits=false, axis on top, xtick={1,3,10,30}, ytick={0.1,1,10}, log ticks with fixed point, colormap name=hamcor, point meta min=0, point meta max=1,
  xlabel={$R\;[r_g]$}, ylabel={$z\;[r_g]$}, title style={font=\scriptsize, yshift=-1mm},
  tick label style={font=\scriptsize}, label style={font=\small}]
\nextgroupplot[title={Mrk 335 (2006)}]
\addplot[matrix plot*, mesh/cols=10, point meta=explicit] table[x=x, y=y, meta=c] {figures/data/mrk335_2006/map_kerr.dat};
\nextgroupplot[title={Mrk 335 (2009)}]
\addplot[matrix plot*, mesh/cols=10, point meta=explicit] table[x=x, y=y, meta=c] {figures/data/mrk335_2009/map_kerr.dat};
\nextgroupplot[title={Mrk 335 (2015)}]
\addplot[matrix plot*, mesh/cols=10, point meta=explicit] table[x=x, y=y, meta=c] {figures/data/mrk335_2015/map_kerr.dat};
\nextgroupplot[title={Mrk 335 (2018)}]
\addplot[matrix plot*, mesh/cols=10, point meta=explicit] table[x=x, y=y, meta=c] {figures/data/mrk335_2018/map_kerr.dat};
\nextgroupplot[title={Mrk 335 (2019)}]
\addplot[matrix plot*, mesh/cols=10, point meta=explicit] table[x=x, y=y, meta=c] {figures/data/mrk335_2019/map_kerr.dat};
\nextgroupplot[title={1H 0707$-$495}]
\addplot[matrix plot*, mesh/cols=10, point meta=explicit] table[x=x, y=y, meta=c] {figures/data/1h0707/map_kerr.dat};
\nextgroupplot[title={IRAS 13224$-$3809}]
\addplot[matrix plot*, mesh/cols=10, point meta=explicit] table[x=x, y=y, meta=c] {figures/data/iras13224/map_kerr.dat};
\nextgroupplot[title={MCG$-$6$-$30$-$15}, colorbar right, colorbar style={width=2mm, ytick={0,0.5,1}, ylabel={$\sum_\phi w_i/\max$}, ylabel style={font=\scriptsize}, tick label style={font=\scriptsize}}]
\addplot[matrix plot*, mesh/cols=10, point meta=explicit] table[x=x, y=y, meta=c] {figures/data/mcg6/map_kerr.dat};
\end{groupplot}
\end{tikzpicture}
\caption{Bootstrap-mean emissivity maps for $a=0.998$ (Kerr transfer, grid extended to $R=1\,r_g$, disc from the ISCO at $1.24\,r_g$). All observations except IRAS~13224$-$3809 are fitted to lags and illumination fraction.}
\label{fig:kerrmaps}
\end{figure*}
\begin{figure}
\centering
\begin{tikzpicture}
\begin{groupplot}[group style={group size=1 by 3, vertical sep=3mm, xlabels at=edge bottom, xticklabels at=edge bottom}, width=0.95\columnwidth, height=0.36\columnwidth,
  xmin=2005, xmax=2020, xtick={2006,2009,2012,2015,2018},
  /pgf/number format/1000 sep={}, xlabel={epoch}, tick label style={font=\scriptsize},
  label style={font=\small}]
\nextgroupplot[ylabel={$\tau_{\rm band}$ [$r_g/c$]}, extra y ticks={0}, extra y tick labels={}, extra y tick style={grid=major, major grid style={densely dotted, gray}}]
\addplot[only marks, mark=*, black, error bars/.cd, y dir=both, y explicit] table[x=year, y=tau, y error=stau] {figures/data/mrk335_epochs.dat};
\nextgroupplot[ylabel={$R_c$ [$r_g$]}, ymin=0, ymax=16]
\addplot[only marks, mark=square*, blue!70!black, error bars/.cd, y dir=both, y explicit] table[x=year, y=Rc, y error plus=Rhi, y error minus=Rlo] {figures/data/mrk335_epochs.dat};
\nextgroupplot[ylabel={$z_c$ [$r_g$]}, ymin=0, ymax=12]
\addplot[only marks, mark=triangle*, red!70!black, error bars/.cd, y dir=both, y explicit] table[x=year, y=zc, y error plus=zhi, y error minus=zlo] {figures/data/mrk335_epochs.dat};
\end{groupplot}
\end{tikzpicture}
\caption{Mrk~335 across five \textit{XMM-Newton} epochs. Top: band-averaged lag in 2--7$\times10^{-4}$\,Hz; middle and bottom: bootstrap median and 16--84 percentiles of the emissivity centroid. Only the 2009 fit uses the illumination constraint (Section~\ref{sec:illum-limit}).}
\label{fig:mrk335}
\end{figure}

\end{appendix}
\end{document}